\documentclass[fleqn,3p,sort&compress]{elsarticle}

\usepackage[breaklinks]{hyperref}
\usepackage[latin1]{inputenc}
\usepackage{graphicx}
\usepackage{color}
\usepackage{amsfonts,amsthm,amssymb}
\usepackage{bm,bbm}
\usepackage[OT2,OT1]{fontenc}
\newcommand\cyr{%
\renewcommand\rmdefault{wncyr}%
\renewcommand\sfdefault{wncyss}%
\renewcommand\encodingdefault{OT2}%
\normalfont
\selectfont}
\DeclareTextFontCommand{\textcyr}{\cyr}

\def\beq{\begin{equation}}
\def\eeq{\end{equation}}
\newcommand{\be}{\begin{eqnarray}}
\newcommand{\ee}{\end{eqnarray}}

\renewcommand{\texttt}{{}}

\def\bs{\begin{subequations}}
\def\es{\end{subequations}}

\newcommand{\tia}[1]{}

\begin{document}

\begin{frontmatter}

\title{
{\bf Super-renormalizable \& Finite Gravitational Theories 
}} 
\author{Leonardo Modesto} 
\author{Les\l aw Rachwa\l{}}

\address{
{\small Department of Physics \& Center for Field Theory and Particle Physics,} \\
{\small Fudan University, 200433 Shanghai, China}
}

\date{\small\today}

\begin{abstract} \noindent
We hereby introduce and extensively study a class of non-polynomial higher derivative theories of gravity 
that realize a ultraviolet (UV) completion of Einstein general relativity. 
These theories are unitary (ghost free) and at most only one-loop divergences survive. The outcome is a class of theories super-renormalizable in even dimension and finite in odd dimension. Moreover, we explicitly prove in $D=4$ that there exists an extension of the theory that is completely finite and all the beta functions vanish even at one-loop. These results can be easily extended in extra dimensions and it is likely that the higher dimensional theory can be made finite too. Therefore we have the  possibility for ``finite quantum gravity" in any dimension.

\end{abstract}


\end{frontmatter}

\section{Introduction}
Quantum abelian and non-abelian gauge theories 
as the most complete embodiment  of particle physics
are all compatible with two guiding principles: 
``{renormalizability" and ``{perturbative theory}" in the quantum field theory framework. 
This is the achievement of a consistent quantum field theory for all but one fundamental interactions.
Indeed, gravity seems to elude so far these patterns
and many authors suggested ingenious solutions to one of the biggest puzzles of our days, but none is completely satisfactory.
The major obstacle, when we try to interface gravity and quantum mechanics is that
Einstein's dynamics is ``non-renormalizable", but in principle there is nothing inconsistent between them. 
Just like for Fermi theory of weak interactions, quantum Einstein's gravity is solid and calculable in the effective field theory framework. The cutoff scale is naturally given for it by Planck energy. 
On the other hand, when the theory is made renormalizable by adding higher derivative operators, 
it is no more unitary and shows up propagation of  ghost states. 
In the end there is a strong tension between renormalizability and 
unitarity in gravitational theories. The key ingredient to overcome this problem
is to introduce a non-polynomial (or non-local) ``kinetic" extension of Einstein's gravity. 
We here use the terminology ``{kinetic part}" for operators linear or quadratic in the gravitational curvature, and 
``potential" for a finite sum of all other local operators in the action.

It is clear from the discussion above that we regard as crucial to find a ``{new theory of gravity}", 
which is unitary and renormalizable or even finite at quantum level.  Moreover we require that such theory is free of singularities at the classical level.
We indeed believe in a one to one correspondence between singularities in classical theory and quantum divergences. 

The aim of this work is to extend {\em classical} Einstein-Hilbert theory 
to make gravity compatible with the above guiding principles (renormalizability and perturbative theory) 
in the ``quantum field theory framework". 
We start with a new unitary non-polynomial higher derivative theory for gravity in a multidimensional spacetime
\cite{modesto, Briscese:2013lna, Krasnikov, Tombo, BM, Chialva, M2, M3, M4, Khoury:2006fg, modestoFinite,Mtheory, AnselmiW} (see also \cite{ModestoMoffatNico, calcagniNL, E1, E2, E3, E4, E5,
Moffat1, Moffat2, Moffat3, corni1}). 
Next we show that it is possible to restrict to a subclass of theories, in which at quantum level only one loop divergences survive. 
Moreover, in such theories these one-loop divergences can be removed by introducing, for example in $D=4$, 
 extra operators that are cubic or quartic in the curvature, typically of the form $O({\cal R}^2 \Box^{\gamma -1} {\cal R})$,
$O({\cal R}^2 \Box^{\gamma -2} {\cal R}^2)$. 
We end up with a completely
finite theory of  quantum gravity, because all the beta functions can be consistently made to vanish by choosing proper coefficients for specially added operators.  
The result can be extended in any dimension and for a more complicated 
curvature potential. In this paper we systematically 
complete the previous work on polynomial \cite{shapiro, HigherDG0} and non-polynomial 
super-renormalizable quantum gravity \cite{modesto, Briscese:2013lna, Krasnikov, Tombo, BM, Chialva, M2, M3, M4, Khoury:2006fg, modestoFinite, Mtheory}.
Our work is also inspired by numerous works on nonlocal infrared modifications of gravity 
\cite{odi, Deser1, Deser2, Modesto:2013jea,Maggiore, MazumdarIR, BiswasSM, Satz:2010uu, Reuter:2003yb}.

{\em Definitions ---} 
The metric tensor $g_{\mu \nu}$ has 
signature $(- + \dots +)$ and the curvature tensors are defined as follows: 
$R^{\mu}_{\nu \rho \sigma} = - \partial_{\sigma} \Gamma^{\mu}_{\nu \rho} + \dots $, 
$R_{\mu \nu} = R^{\rho}_{\mu  \rho \nu}$,  
$R = g^{\mu \nu} R_{\mu \nu}$. With symbol ${\cal R}$ we generally denote one of the above curvature tensors.

\section{Modern Gravity}
In this section we introduce a 
 ``New Gravity" theory
in a $D$-dimensional spacetime assuming the following consistency requirements: 
\begin{enumerate}
\renewcommand{\theenumi}{\arabic{enumi}}
\item Unitarity. A general theory is well defined, if 
the corresponding propagator has only first poles 
with real masses (no tachyons) and with positive
residues (no ghosts).
\item Super-renormalizability or Finiteness. This
hypothesis makes consistent the theory at quantum level
on the same footing as for all the other fundamental interactions.
\item Lorentz invariance. 
This is a symmetry of
nature 
well tested experimentally beyond the Planck mass.
\item  
The classical energy conditions can only  
be violated, because higher-derivative
operators are present in the classical theory. This property is crucial to avoid singularities, that plague almost all the solutions of Einstein's gravity 
\cite{ModestoMoffatNico, calcagnimodesto, Cnl1,Cnl2,BambiMalaModesto, BambiMalaModesto2,koshe1,koshe2,koshe3,koshe4, kosheN}. 
\end{enumerate}

The most general multidimensional theory compatible with the above requirements consists of a non-polynomial 
(or nonlocal) sector and a local curvature potential, namely 
\be
&& \hspace{-0.4cm}
\mathcal{L} = - 2 \, \kappa_D^{-2}\sqrt{|g|} \left( R + 
G_{\mu \nu}    
\frac{ e^{H(-\Box_{\Lambda})} -1}{\Box} R^{\mu \nu} +
 \mathcal{V}   \right)  , 
 \nonumber   \\
 && \hspace{-0.4cm}
 {\rm with} \hspace{0.2cm}\mathcal{V} = 
 \!\!\!
 \sum_{n=3}^{\gamma + \rm{N} + 2}  \!\!\! \alpha_{2 n} \, \Lambda^{2 - 2 n} 
 \, \mathcal{O}_{2 n}  (\partial_{\rho } g_{\mu \nu }) \, , 
  \label{Action0}
 \ee 
 where $\Lambda$ is an invariant mass scale in our fundamental theory, 
$\mathcal{O}_{2n} (\partial_{\rho} g_{\mu \nu})$ denotes schematically all the generally covariant scalar terms $O({\cal R}^3)$ containing 
``$2 n \geqslant 6$" derivatives of the metric tensor $g_{\mu \nu}$. 
Using a schematic notation we can classify the operators $O({\cal R}^3)$ as follows, 
\be
&& \hspace{-1cm} 
O_{6}=\left\{ {\cal R}^{3}\right\} , \nonumber \\
&&\hspace{-1cm} 
O_{8}=\left\{ {\cal R}^{4},\nabla^{2}{\cal R}^{3}\right\} , \nonumber \\
&&\hspace{-1cm} 
O_{10}=\left\{ {\cal R}^{5},\nabla^{2}{\cal R}^{4},\nabla^{4}{\cal R}^{3}\right\} , \nonumber \\
&&
\ldots \nonumber \\
&& \hspace{-1cm} 
O_{2\gamma+2 {\rm N}+4}=\left\{ {\cal R}^{\gamma+{\rm N}+2},\nabla^{2}{\cal R}^{\gamma+{\rm N}+1},\nabla^{4}{\cal R}^{\gamma+{\rm N}},
\ldots,\nabla^{2\gamma+2 {\rm N}-4}{\cal R}^{4},\nabla^{2\gamma+2 {\rm N}-2}{\cal R}^{3}\right\}  \, , 
\label{operators}
\ee
where indices and tensorial structure have been neglected.

For the specific case of a finite theory it is sufficient to concentrate on the following reduced 
potentials in even dimension (in odd dimension we do not need to introduce any potential to make the theory finite), 
\be
 && \hspace{-0.4cm}
 \mathcal{V}
 = \sum_{k=4}^{{\rm N}+2} \sum_i s_{k,i} \, \left( \nabla^{2( \gamma + {\rm N} + 2 - k)} \, {\cal R}^k \right)_i  \, 
  \, \, , \label{K0}
 \ee
 where the sum must include at least the minimal set of operators (with different tensorial structure), which we need to make the theory finite. From (\ref{Action0}) $s_i \equiv \tilde s_i \Lambda^{-2\gamma -2{\rm N} -2}$ are dimensionful parameters in these operators with the highest possible number of derivatives. Moreover  
$\Box = g^{\mu\nu} \nabla_{\mu} \nabla_{\nu}$ is the covariant box operator, 
$G_{\mu\nu}$ is the Einstein tensor, the integer parameter $\gamma$ and the entire function 
$H(-\Box_{\Lambda})$ will be shortly defined. The capital $\rm{N}$ is defined to be the following function of the spacetime dimension $D$:
$2 \mathrm{N} + 4 = D_{\rm odd} +1$ in odd dimensions and $2 \mathrm{N} + 4 = D_{\rm even}$
in even dimensions  in order to avoid fractional powers of the d'Alembertian operator.
Finally, the entire function $V^{-1}(z) \equiv \exp H(z)$ ($z \equiv - \Box_{\Lambda} \equiv - \Box/\Lambda^2$)
satisfies the following general conditions \cite{Tombo}:
\begin{enumerate}
\renewcommand{\theenumi}{(\roman{enumi})}
\item 
$V^{-1}(z)$ is real and positive on the real axis and it has no zeros on the 
whole complex plane $|z| < + \infty$. This requirement implies that there are no 
gauge-invariant poles other than the transverse massless physical graviton pole;
\item
$|V^{-1}(z)|$ has the same asymptotic behavior along the real axis at $\pm \infty$; 
\item 
There exists $\Theta>0$ and $\Theta<\pi/2$, such that asymptotically
\be
&& 
|V^{-1}(z)| \rightarrow | z |^{\gamma + \mathrm{N}+1},\,\, {\rm when }\,\, |z|\rightarrow + \infty\,\, {\rm with}\nonumber \\ 
&& \gamma\geqslant D_{\rm even}/2 
\,\,\,\, {\rm and} 
\,\,\, \, 
 \gamma\geqslant (D_{\rm odd}-1)/2 \,\,\, {\rm respectively} , 
\label{tombocond}
\ee 
for the complex values of $z$ in the conical regions $C$ defined by: 
$$C = \{ z \, | \,\, - \Theta < {\rm arg} z < + \Theta \, ,  
\,\,  \pi - \Theta < {\rm arg} z < \pi + \Theta\}.$$
\end{enumerate}
The last condition is necessary to achieve the maximum convergence of the theory in
the UV regime.  
The necessary asymptotic behavior is imposed not only on the real axis, but also on the conical regions, that surround it.  
In an Euclidean spacetime, the condition (ii) is not strictly necessary if (iii) applies.

In $D=4$ the minimal theory compatible with the properties (i)-(iii) and finite at the quantum level 
contains only two local extra vertices, namely
\be
&&  \hspace{-0.5cm}
\hspace{0cm}\mathcal{L}_{\rm g} = - 2 \kappa_D^{-2}  \sqrt{|g|} \Big( R  
+ G_{\mu \nu} \,  \frac{ e^{H(-\Box_{\Lambda})} -1}{\Box}   R^{\mu \nu} 
+  s_{1}  R^2 \, \Box^{ \gamma -2} R^2 
+ s_{ 2} 
R_{\mu \nu} R^{\mu\nu} \, \Box^{ \gamma -2} R_{\rho \sigma} R^{\rho \sigma} \Big) .
\label{Action2}
\ee

An explicit example of $\exp H(z)$, that has the properties (i)-(iii) can be easily constructed \cite{Tombo}, 
\be
&& \hspace{-0.5cm}
V^{-1}(z) \equiv e^{H(z)}=\exp\left({\sum_{n =1}^{+ \infty} \, \frac{(-1 )^{n+1} \, p(z)^{2 n}}{2n  \, n!}}\right)
\label{VlimitB}
= e^{\frac{1}{2} \left[ \Gamma \left(0, p(z)^2 \right)+\gamma_E  + \log \left( p(z)^2 \right) \right] }=\\
&&  \hspace{0.7cm}
= e^{\frac{1}{2} \left[ \Gamma \left(0, p(z)^2 \right)+\gamma_E  \right] } \, \left| p(z) \right|  \label{VlimitC}
=  \underbrace{
e^{\frac{\gamma_E}{2}} \,
\left| p(z) \right| 
}_{V^{-1}_{\infty}(z) } +
    \underbrace{ 
\left(   e^{\frac{1}{2}  \Gamma \left(0, p(z)^2 \right)  }  -1 \right) 
e^{\frac{\gamma_E}{2}} 
\, \left| p(z) \right|   
}_{V^{-1}(z) -V^{-1}_{\infty}(z) }  , 
\label{Vlimit1}
\ee
where the equality between (\ref{VlimitB}) and (\ref{VlimitC}) is correct only on the real axis. 
The  polynomial $p(z)$ of degree $\gamma +\mathrm{N}+1$ is such that $p(0)=0$, which gives the correct low energy limit of our theory. 
In (\ref{VlimitB}), (\ref{Vlimit1})
 $\gamma_E \approx 0.577216$ is the Euler-Mascheroni constant and 
$
\Gamma(0,z) = \int_z^{+ \infty}  d t \, e^{-t} /t 
$ 
is the incomplete gamma function with its first argument vanishing. 
The angle $\Theta$ defining 
 cones $C$ turns out to be $ \pi/(4 (\gamma + \mathrm{N} + 1))$. 
A crucial property of the form factor for the convergence of the theory in UV is that on the real axis 
\be
&& \hspace{0.0cm}  V^{-1}(z) \rightarrow V^{-1}_{\infty}(z) = e^{\frac{\gamma_E}{2}} \, |p(z)|, 
\,\,\,\, {\rm when} \,\,\, |z| \rightarrow +\infty 
 \nonumber \\  
&& {\rm and} 
\,\,\,\, \,\,\,
 \lim_{|z| \rightarrow +\infty} 
\left(\frac{V^{-1}(z)}{e^{\frac{\gamma_E}{2}} |p(z)| } - 1 \right) z^n = 0
\,\,\,\, \forall \, n \in \mathbb{N}\, .
\label{property}
\ee
This can be easily justified by expanding to the next to leading order for large $z$ (or equivalently for large values of the polynomial $p(z)$). The form factor on the real axis reads:
\be
&& \hspace{-1cm}
V^{-1}(z)  
= e^{e^{-p(z)^2} \left(  \frac{1}{2 p(z)^2} - \frac{1}{2 p(z)^4} + O\left( \frac{1}{p(z)^6} \right)  \right)  } 
\, e^{\frac{\gamma_E}{2}}\, \left| p(z) \right|  \, , \\
&& \hspace{-1cm}
V^{-1}(z) - V^{-1}_{\infty}(z)  
=  \left( e^{-p(z)^2} \left(  \frac{1}{2 p(z)^2} - \frac{1}{2 p(z)^4} + O\left( \frac{1}{p(z)^6} \right)   \right) + O \left(e^{-2p(z)^2} \right)  \right)  
\, e^{\frac{\gamma_E}{2}}\, \left| p(z) \right|  \, , 
\label{LimitVP}\\
&& \hspace{-1cm}
\lim_{|z| \rightarrow + \infty} e^{\frac{1}{2}\Gamma \left(0, p(z)^2  \right)} =1 \,, \quad {\rm because} \quad p(z)^2 \to +\infty\, \quad {\rm when }\quad |z| \rightarrow + \infty\,. 
\ee

\subsection*{Propagator}
\label{gravitonpropagator}
Splitting the spacetime metric into the flat Minkowski background $\eta_{\mu\nu}$ and the fluctuation $h_{\mu \nu}$ 
defined by $g_{\mu \nu} =  \eta_{\mu\nu} + \kappa_D \, h_{\mu\nu}$, 
we can expand the Lagrangian (\ref{Action0}) to the second order in $h_{\mu\nu}$.
The result of this expansion together with a gauge fixing term $\mathcal{L}_{\rm GF}$ reads \cite{HigherDG}:
\be
\mathcal{L}_{\rm quad} + \mathcal{L}_{\rm GF} = \frac{1}{2} h^{\mu\nu} \mathcal{O}_{\mu\nu, \rho\sigma} \, h^{\rho\sigma} \, ,
\label{O}
\ee
where the kinetic operator $\mathcal{O}$ is made of two terms. 
The first one comes from the quadratic expansion of (\ref{Action2})
and the other one from the following usual harmonic gauge-fixing term
$\mathcal{L}_{\rm GF}  = \xi^{-1}  \partial^{\nu}h_{\nu \mu } \omega(-\Box_{\Lambda}) \partial_{\rho}h^{\rho \mu}$, where
$\omega( - \Box_{\Lambda})$ is a weight functional \cite{Stelle, Shapirobook}.
Obviously the d'Alembertian operator in $\mathcal{L}_{\rm quad}$ and in the weight $\omega$ must be conceived on the flat spacetime. 
Inverting the operator $\mathcal{O}$ \cite{HigherDG}, we find the  
two-point function in the harmonic gauge ($\partial^{\mu} h_{\mu\nu} = 0$),
\be
&& \hspace{-1cm}   \mathcal{O}^{-1} \!=\!
 \frac{V(  k^2/\Lambda^2 )  } {k^2} \!
\left( P^{(2)} 
- \frac{P^{(0)}}{D-2 }  \right)  + 
 \frac{\xi (2P^{(1)} + \bar{P}^{(0)} ) }{2 k^2 \, \omega( k^2/\Lambda^2)} .
 \label{propagator}
\ee
The tensorial 
indices for the operator $\mathcal{O}^{-1}$ and the projectors $\{ P^{(0)},P^{(2)},P^{(1)},\bar{P}^{(0)}\}$ have been omitted. The above projectors are defined by  
\cite{HigherDG, VN}:  
\be
 && \hspace{-0.2cm}
 P^{(2)}_{\mu \nu, \rho \sigma}(k) = \frac{1}{2} ( \theta_{\mu \rho} \theta_{\nu \sigma} +
 \theta_{\mu \sigma} \theta_{\nu \rho} ) - \frac{1}{D-1}  \theta_{\mu \nu} \theta_{\rho \sigma} ,
 \nonumber
 \\
 && \hspace{-0.2cm}
P^{(1)}_{\mu \nu, \rho \sigma}(k) = \frac{1}{2} \left( \theta_{\mu \rho} \omega_{\nu \sigma} +
 \theta_{\mu \sigma} \omega_{\nu \rho}  +
 \theta_{\nu \rho} \omega_{\mu \sigma}  +
  \theta_{\nu \sigma} \omega_{\mu \rho}  \right) , \nonumber   \\
   &&
  \hspace{-0.2cm}
P^{(0)} _{\mu\nu, \rho\sigma} (k) = \frac{1}{D-1} \theta_{\mu \nu} \theta_{\rho \sigma}  , \,\,\,\, 
\bar{P}^{(0)} _{\mu\nu, \rho\sigma} (k) =  \omega_{\mu \nu} \omega_{\rho \sigma}, 
 \label{proje2}
\ee
where
 $\theta_{\mu \nu} = \eta_{\mu \nu} - k_{\mu } k_{\nu }/k^2$ and $\omega_{\mu \nu } = k_{\mu} k_{\nu}/k^2$.
 
 The tensorial structure in(\ref{propagator}) is the same of Einstein gravity, but the multiplicative 
 form factor $V(-\Box_{\Lambda})$ makes the theory strongly convergent without the need to modify the spectrum
 or introducing instabilities.

\section{Strict analysis of quantum divergences}

Let us then examine the UV behavior of the quantum theory and what operators in the action
are source of divergences.
Assuming the form factor to be asymptotically polynomial (\ref{Vlimit1}), the most general multidimensional Lagrangian density (\ref{Action0}) reads\footnote{
In $D=4$ the Lagrangian density reads:
\be
  \hspace{-0.5cm}
\hspace{0cm}\mathcal{L}_{\rm g} =\bar{\lambda} - \frac{2}{\kappa_4^{2}}  R 
-  2 \, G_{\mu \nu} \,  \frac{ e^{H(-\Box_{\Lambda})} -1}{ \tilde{\kappa}_4^2 \Box}   R^{\mu \nu}  
+ (a_0 - \tilde{a}_0)R^2 
\label{Action2b}
+ ( b_0 - \tilde{b}_0) R_{\mu \nu}^2  
-  \frac{2 s_1}{ \tilde{\kappa}_4^2} \, R^2 \, \Box^{ \gamma -2} R^2 - \frac{2 s_2}{ \tilde{\kappa}_4^2}
R_{\mu \nu} R^{\mu\nu} \, \Box^{ \gamma -2} R_{\rho \sigma} R^{\rho \sigma}  .
\nonumber
\ee
}, 
\be
&& \hspace{-0.2cm}
\mathcal{L}_{\rm g} = 
\mathcal{L}_{{\rm Kinetic} } -  2 \tilde{\kappa}_D^{-2} \, \mathcal{V}   + \bar{\lambda}\,, \label{mostgeneral}\\
&& \hspace{-0.2cm}
\mathcal{L}_{{\rm Kinetic} } \equiv 
- \frac{2 }{\kappa_D^{2} } R - 
2 \, G_{\mu \nu}    
\frac{ e^{H(-\Box_{\Lambda})} -1}{\tilde{\kappa}_D^2\Box}
 R^{\mu \nu}  + \mathcal{L}_Q \, , 
 \nonumber  \\
&&  \hspace{-0.2cm}
 \mathcal{L}_Q =  \sum_{n=0}^{\mathrm{N}} 
 \Big[ (a_n - \tilde{a}_n) R \, \Box^n \, R  
  + (b_n -  \tilde{b}_n)  R_{\mu \nu} \, \Box^n \, R^{\mu \nu} \Big] ,
 \nonumber   \\
 && \hspace{-0.2cm}
\mathcal{V} \equiv  - \frac{\tilde{\kappa}_D^2}{2} V_{<} + V_{><} + V_{K}  , \nonumber \\
 && \hspace{-0.2cm}
 V_{<} \equiv 
 \sum_i c_{3,i}^{(3)} \left( {\cal R}^3 \right)_i + \dots + \sum_{k=3}^{{\rm N}+2} \sum_i c_{k,i}^{(\rm{N}+2)} \left( \nabla^{2({\rm N}+2-k)} {\cal R}^{k} \right)_i = \sum_{j=3}^{{\rm N}+2} \sum_{k=3}^{j} \sum_i c_{k,i}^{(j)} \left( \nabla^{2(j-k)} {\cal R}^k \right)_i\, , \nonumber \\ 
 && \hspace{-0.2cm}
 V_{><} \equiv 
\sum_{k=3}^{{\rm N}+3} \sum_i d_{k,i}^{({\rm N} + 3)} \left(\nabla^{2({\rm N}+3-k)} {\cal R}^k \right)_i + \dots + \sum_{k=3}^{\gamma + {\rm N}+1} \sum_i d_{k,i}^{( \gamma + {\rm N} +1)} \left(\nabla^{2(\gamma + {\rm N}+1-k)} {\cal R}^k \right)_i= \nonumber \\ 
 && \hspace{0.6cm}
= \sum_{j={\rm N}+3}^{\gamma+{\rm N}+1} \sum_{k=3}^{j} \sum_i d_{k,i}^{(j)} \left(\nabla^{2(j-k)} {\cal R}^k \right)_i\,, \nonumber \\ 
&& \hspace{-0.2cm}
 V_{K,\,{\rm general}} = \sum_{k=3}^{\gamma +{\rm N}+2} \sum_i s_{k,i} \, \left(  \nabla^{2 (\gamma + {\rm N}+2 -k )} \, {\cal R}^k \right)_i \,\, , 
  \nonumber 
 \ee 
 where we also introduced all possible local quadratic terms in the curvature up to $2 {\rm N} +4$
 derivatives.
As will be clear shortly only the following coupling constants are subject to renormalization
\be
\alpha_i \equiv \{ \bar{\lambda}, \kappa_D^{-2}, a_n, b_n, {c}_{k,i}^{(3)}, \dots, {c}_{k,i}^{(\rm{N} + 2)} \} \, .
\label{coupling}
\ee

At classical level we choose the following identification
\be
\alpha_i = {\rm const} =\{ \tilde{\bar{\lambda}}, \tilde{\kappa}_D^{-2}, \tilde{a}_n, \tilde{b}_n, \tilde{c}_{k,i}^{(3)}, \dots, \tilde{c}_{k,i}^{(\rm{N} + 2)}  \} 
\label{CIde}
\ee
and the action (\ref{mostgeneral}) reduces to the unitary theory (\ref{Action0}). 

At quantum level we face with two possibilities. If the theory is finite all the beta functions vanish, we have scale invariance and 
the classical identification (\ref{CIde}) is valid also at the quantum level. 
If the theory is renormalizable, then  the parameters $\tilde{\bar{\lambda}}$, 
$\tilde{\kappa}_D^{-2}$, $\tilde{a}_n$, $\tilde{b}_n$ and $\tilde{c}_{k,i}^{(j)}$ in (\ref{mostgeneral}) are just the initial conditions 
for renormalization group equations of the running coupling constants $\bar{\lambda}$, 
$\kappa_D^{-2}$, $a_n$, $b_n$, ${c}_{k,i}^{(j)}$. 
The operators $R, \mathcal{L}_Q, V_<$ will be multiplied by 
the logarithm of  the energy scale $\mu$ coming from the running of 
all the coupling constants $\alpha_i(\mu)$. However, these contributions can be absorbed 
in the finite parts of the one loop effective action, which involve the same operators $R, \mathcal{L}_Q, V_<$ with $\log (- \Box/\mu^2)$  in between. This a consequence of the renormalization group invariance  
as  we will show explicitly at the end of this section. 

For the coupling constants $c_{k,i}^{(j)}$, the lower index ``$i$" runs over all possible operators with a fixed power of curvature and fixed  number of derivatives on the metric enumerated by 
$2j \in [6, 2\rm{N} +4]$. 
For the constant parameters $d_{k,i}^{(j)}$, the lower index ``$i$" labels similarly all possible operators with a  number of derivatives on the metric in  the range 
$2j \in [2\rm{N} +6, 2 \gamma + 2 {\rm N} +2]$. Index ``$k$" counts the overall power of covariant curvature in a term. All the operators in $V_{<}$ and $V_{><}$ are at least cubic in curvature.

In the high energy regime, 
the graviton propagator in momentum space 
schematically scales as 
\be
\mathcal{O}^{-1}(k) \sim \frac{1}{k^{2 \gamma +2 \mathrm{N} +4}} \,\,\,\,\,\,
\mbox{in the UV} \, . 
\label{OV} 
\ee
The vertices can be collected in four different sets, that may involve 
or not the entire functions $\exp H(z)$. 
In what follows we omit the tensor indices to make the analysis slender. 
The first set comes from the operators in $V_{<}$ and $\mathcal{L}_Q$,
\be
 \hspace{-0.8cm} 
{\rm set} \, 1 :   \,\, 
 {\cal R}, \, {\cal R}^2 , \, {\cal R}^3 , {\cal R} \, \Box {\cal R}, \dots , \, {\cal R}^{\mathrm{N}+2} , {\cal R} \, \Box^{\rm N} {\cal R} \,\,\,   \label{Vertex1} 
\,\,\, \Longrightarrow \,\,\,\,
h^m (\partial^2 h) , \, h^m (\partial^2 h)^2 , \,
h^m (\partial^2 h)^3 , \dots , h^m (\partial^2 h)^{\mathrm{N}+2}   .  
\ee
The above operators can not give origin to divergences for the integer $\gamma > D/2$.
The second set derives from the form factor $\exp H$, namely it contains the operators involving
\be 
{\rm set}\, 2 :  \,\,  {\cal R}\, \frac{ \exp H( - \Box_{\Lambda}) }{\Box} \, {\cal R} \,\,\,  
\label{Vertex2} 
\Longrightarrow
\,\,\,\, 
h^m \, (\partial^2 h)
\,   \frac{p(-\Box_{\Lambda})}{\Box} \,(\partial^2 h) \, .
\ee
These operators for sure give contribution to the divergences, because they scale like the propagator. 
The third set originates from the operators involved in 
the potential $V_K$, 
\be 
{\rm set} \, 3  :  \,\, {\cal R}^{\frac{D}{2} } {\cal R} \,  \nabla^{2 \gamma - 4 } \, {\cal R}^2   \,\,\, 
\label{Vertex3} 
\Longrightarrow
\,\,\,\, h^m \, (\partial^2 h)^{\frac{D}{2} }
\, \nabla^{2 \gamma - 4 } \,
(\partial^2 h)^{2}\, .
\ee
Even in this case we have non zero contribution to the divergences.

The last set comes from the potential 
$V_{><}$
\be &&
{\rm set} \, 4  :   \,\, {\cal R}^{{\rm N} +3}    ,  \dots  \,  , {\cal R}^{\gamma + {\rm N} } 
\, , {\cal R}^{\gamma + {\rm N} +1} \,
   {\cal R}^{ {\rm N}+2} \Box^{\gamma - 3} {\cal R} 
 , \,  {\cal R}^{ {\rm N}+2} \Box^{\gamma - 2} {\cal R}
\nonumber \\
&& \hspace{1cm} \Longrightarrow   
h^m  (\partial^2 h)^{\rm{N}+2} 
\, \Box^{\gamma - 3 } \,
(\partial^2 h)
\, , \, 
h^m  (\partial^2 h)^{\rm{N}+2} 
\, \Box^{\gamma - 2 } \,
(\partial^2 h) \, .
\label{Vertex4}
\ee
The subset of operators $O\left( (\partial_\rho g_{\mu\nu})^{2 \gamma} \right)$ in $V_{><}$  
can also contribute to the divergences (see the last two operators in (\ref{Vertex4})). 
In (\ref{Vertex1})-(\ref{Vertex4}) the exponent ``$m$"  
comes from the operators expansion in the graviton field.

From the propagator (\ref{OV}) and the vertices (\ref{Vertex1})-(\ref{Vertex4}), 
an upper bound on the superficial degree of divergence 
in a spacetime of even or odd dimension reads 
\be 
&&
\omega(G)_{\rm even} =  D_{\rm even} - 2 \gamma  (L - 1)    \, , \,\,\,\,
\label{even}\\
&& 
\omega(G)_{\rm odd} = D_{\rm odd} - (2 \gamma+1)  (L - 1).
\label{odd}
\ee
In 
(\ref{odd}) we used the topological relation between vertices $V$, internal lines $I$ and 
number of loops $L$: $I = V + L -1$. 
Thus, if $\gamma > D_{\rm even}/2$ or $\gamma > (D_{\rm odd}-1)/2$, in the theory only 1-loop divergences survive.  
Therefore, 
the theory is super-renormalizable \cite{Krasnikov, E1, E2, E3, E4, E5}
and only a finite number of coupling constants is renormalized in the action (\ref{mostgeneral}), i.e. 
$\kappa_D^{-2}$, $\bar{\lambda}$, $a_n$, $b_n$ together with the finite number of couplings 
in the potential $V_{<}$. 

Let us now expand on the one-loop divergences for the case 
$p(z) = z^{\gamma +{\mathrm N} + 1}$. 
The main divergent integrals contributing to the one-loop effective action have the following form
\be 
\hspace{-0.4cm}
\int \!\! \frac{d^D k}{(2 \pi)^D}  \left\{ \prod_{i=1}^{s} 
\frac{1}{(k+p_i)^{2 n}} \right\} \! P_{2 s  n}(k)  . 
\label{integral}
\ee
$P_{2sn}(k)$ is a polynomial function of degree $2 sn$ in the momentum $k$ 
(generally it also depends on the external momenta $\bar{p}_a$),  
$p_i = \sum_{a=1}^i \bar{p}_a$. 
The positive integer $n$ is: $n = \gamma +{\rm N} +2$ for $h_{\mu\nu}$, 
$n = 1$ for the ghosts $C,\bar{C}$ and $n = \gamma +{\rm N} +1$ for the third ghost $b_\alpha$
(the gauge fixing and ghost action will be explicitly defined in the next section.) 
We can write, as usual,
\be
&& \hspace{-0.5cm}
\prod_{i=1}^{s} 
\frac{1}{ (k+p_i )^{2 n}}  
= {\rm c} \!  \int_0^1 \! \left( \prod_{i=1}^s x_i^{n-1} d x_i \!  \right) 
\delta\left( 1 - {\sum_{i=1}^s x_i} \right)
 \frac{1}{[k^{\prime 2} + M^{2}]^{n s}}  \, , \,\,\,\, \nonumber\\
 &&  \hspace{-0.5cm} 
 k^{\prime} = k + \! \sum_{i=1}^s x_i p_i   \, , \,\, \, 
 M^{ 2}= \sum_{i=1}^s p_i^2 x_i 
 - \left( \sum_{i=1}^s x_i p_i \right)^{\!\! 2} \!\!  \, . \nonumber  
\ee
where ${\rm c} = {\rm const}$. 
In (\ref{integral}) we move outside the convergent integrals in $x_i$ and we replace  
$k^{\prime}$ with $k$  
\be
\int \!\! \frac{d^D k}{(2 \pi)^D} 
\frac{P^{\prime}(k, p_i, x_i)_{2 n s} }{( k^2 + M^{ 2})^{n s }} \, .
\label{int2}
\ee
Using Lorentz invariance and neglecting the argument $x_i$, we replace the polynomial 
$P^{\prime}(k, p_i, x_i)_{2 n s}$ with a polynomial of degree $n \times s$ in $k^2$, 
namely $P^{\prime \prime}(k^2, p_i)_{n s}$. 
Therefore the integral (\ref{int2}) reduces to
\be
\int \!\!  \frac{d^D k}{(2 \pi)^D} 
\frac{P^{\prime \prime}(k^2, p_i)_{n s} }{( k^2 + M^{2} )^{n s }} \, .
\label{int3}
\ee
We can decompose the polynomial $P^{\prime \prime}(k^2, p_i)_{n s} $ 
in a product of external and internal momenta in order
to obtain the divergent contributions
\be
&& \hspace{-0.4cm} 
P^{\prime \prime}(k^2, p_i)_{n s} = \sum_{\ell=0}^{[ D/2 ]} \alpha_{\ell}(p_i) k^{2 n s-2 \ell} = 
k^{2 n s} \alpha_0 + k^{2 n s-2} \alpha_1(p_i) + k^{2 n s - 4 } \alpha_2(p_i)   
+ \dots \, . \label{expa}
\ee

\indent Given $p(z) = z^{\gamma +{\mathrm N} + 1}$ and switching off $V_{><}$,
if all the vertices but one come from set 2 in (\ref{Vertex2}) or set 3 in (\ref{Vertex3}), then the integral (\ref{int3}) 
does not give any logarithmic divergence. 
We find logarithmic divergences only when {\em all} the vertices come from set 2 in (\ref{Vertex2}) 
or set 3 in (\ref{Vertex3}) and then 
the contribution   
to the amplitude follows from (\ref{int3}) and the equation (\ref{expa}), namely\be
&&\hspace{-0.5cm}
 \sum_{\ell=0}^{[ D/2 ]} 
\int \!\!  \frac{d^D k}{(2 \pi)^D} 
\frac{ \alpha_{\ell}(p_i) k^{2 n s-2 \ell}  }{( k^2 + M^{2} )^{n s }} 
\label{int4} 
 =  \sum_{\ell=0}^{[ D/2 ]} 
  \frac{i \alpha_{\ell}(p_i) (M^2)^{\frac{D}{2}- \ell }}{(4 \pi)^{\frac{D}{2}}} \, \frac{\Gamma \left(\ell - \frac{D}{2}  \right) 
  \Gamma \left(ns -\ell + \frac{D}{2}  \right)}{\Gamma \left( \frac{D}{2}  \right) \Gamma(n s)}  \,  .
\nonumber 
\ee
The counterterms, all having the same mass dimension, are elements of the following set,
\be
\hspace{-0.7cm}
\left\{  \frac{1}{\epsilon} \, \left( {\cal R}^{\mathrm{N} +2} \right)_i \,  , \,\,\,  \frac{1}{\epsilon} \left(\nabla^2 {\cal R}^{{\rm N}+1} \right)_i \,  , \ldots, 
\,\,\,  \frac{1}{\epsilon}\left( \nabla^{2{\rm N}} {\cal R}^2  \right)_i \right\} = \left\{ \frac{1}{\epsilon}\left( \nabla^{2({\rm N}+2-k)} {\cal R}^k  \right)_i : 2\leq k \leq {\rm N} +2, k \in {\mathbb N} \right\} ,
\label{ellediv}
\ee
where 
$\epsilon=D-4$ is the UV cutoff in dimensional 
regularization. 
The outcome is that, for big enough $\gamma$, we have counterterms only at the order ${\cal R}^{\mathrm{N} +2}$. This observation is a first step in the direction to find a finite quantum theory. 
For example, in $D=4$ the counterterms are $R^2$ and $R_{\mu \nu}^2$, but there are no divergent contributions 
proportional to $R$ or $\bar{\lambda}$ (cosmological constant). 
This is a property of the theory defined   
by the particular polynomial $p(z) = z^{\gamma +{\rm N}+ 1}$ and $V_{><}=0$. 
However, if we assume the more general polynomial
\be
\hspace{-0.3cm}
 p_{\gamma+\mathrm{N}+1}(z) = a_{ \mathrm{N}} \, z^{\gamma + \mathrm{N} + 1} 
+ \dots + a_{\mathrm{N} - \frac{D}{2} } \, z^{\gamma + \mathrm{N}+1 - \frac{D}{2}}  \, 
\label{poly} 
\ee
and/or we switch on $V_{><}$,  
 then the other couplings  
 are also renormalized due to
 counterterms with less derivatives.
 
\subsection*{Renormalization \& asymptotic freedom}
The renormalized Lagrangian in the multiplicative renormalization scheme reads as follows,
\be
&& \hspace{-0.3cm}
\mathcal{L}_{\rm g}^{\rm Ren} = 
\mathcal{L}_{{\rm Kinetic} }^{\rm Ren} - 2 \tilde{\kappa}^{-2}_D \mathcal{V}^{\rm Ren}   +Z_{\bar{\lambda}} \bar{\lambda}\,, \nonumber \\
&& \hspace{-0.3cm}
\mathcal{L}_{{\rm Kinetic} }^{\rm Ren} \equiv 
- \frac{2 Z_{\kappa_D^{-2}} }{\kappa_D^{2} } R -
2 \, G_{\mu \nu}    \frac{ e^{H(-\Box_{\Lambda})} -1}{\tilde{\kappa}_D^2 \Box}
 R^{\mu \nu}  + \mathcal{L}_Q^{\rm Ren} \, , 
 \nonumber  \\
&&  \hspace{-0.3cm}
 \mathcal{L}_Q^{\rm Ren} = 
 \sum_{n=0}^{\mathrm{N}} 
  \Big[ (Z_{a_n} a_n - \tilde{a}_n) R \, \Box^n \, R   
  + ( Z_{b_n} b_n -  \tilde{b}_n)  R_{\mu \nu} \, \Box^n \, R^{\mu \nu} \Big] ,
 \nonumber   \\
 && \hspace{-0.3cm}
 \mathcal{V}^{\rm Ren} \equiv  - \frac{\tilde{\kappa}_D^2}{2} V_{<}^{\rm Ren}
  + V_{><} + V_{K}  \, , \,\,\,\,\, \nonumber \\
 && \hspace{-0.3cm}
 V_{<}^{\rm Ren} = \sum_{j=3}^{{\rm N}+2} \sum_{k=3}^{j} \sum_i  Z_{c_{k,i}^{(j)}} c_{k,i}^{(j)} \left( \nabla^{2(j-k)} {\cal R}^k \right)_i \, , 
 \nonumber \\ 
 && \hspace{-0.3cm}
 V_{><} =\sum_{j={\rm N}+3}^{\gamma+{\rm N}+1} \sum_{k=3}^{j} \sum_i d_{k,i}^{(j)} \left(\nabla^{2(j-k)} {\cal R}^k \right)_i \, , 
 \,\,\,\, 
 \nonumber\\
&& \hspace{-0.3cm}
V_{K} = \sum_{k=4}^{{\rm N}+4} \sum_i s_{k,i} \, \left( \, \nabla^{2 (\gamma +{\rm N}+2-k)} \, {\cal R}^k \right)_i\,,  \label{mostgeneralRen}
 \ee 
while the potentials $V_{><}$ and $V_{K}$ are not subject to renormalization. In the formula above we already wrote a minimal form of the killer potential $V_K$.

We now expand on the renormalization of the Lagrangian in (\ref{mostgeneral}) and the running of the coupling constants.
We start with the classical action written in terms of renormalized couplings and then we add counterterms to subtract divergences.
The counterterms may be displayed by explicitly adding and subtracting the classical action in 
$\mathcal{L}^{\rm Ren}$ (\ref{mostgeneralRen}),
\be 
&&
 \hspace{-1.0cm} 
 \mathcal{L}^{\rm Ren} =  \mathcal{L}_{\rm g} + \mathcal{L}_{\rm ct}   
=  \mathcal{L}_{\rm g} 
 - 2   (Z_{\kappa_D^{-2}} -1)  \kappa_D^{-2} \, R + (Z_{\bar{\lambda}} -1) \bar{\lambda}
\nonumber  
+  
 \sum_{j=3}^{{\rm N}+2} \sum_{k=3}^{j} \sum_i \left(Z_{c_{k,i}^{(j)}} -1\right) c_{k,i}^{(j)} \left( \nabla^{2(j-k)} {\cal R}^k \right)_i 
   \\
&& \hspace{0.1cm}
+ \sum_{n=0}^{\mathrm{N} }  \Big[ 
   ( Z_{a_n} -1)a_n   R  \, \Box^n  R  + ( Z_{b_n} -1)  b_n R_{\mu \nu}  \Box^n  R^{\mu \nu}   \Big] ,  
   \label{actionRen2}
   \ee 
where $\mathcal{L}_{\rm ct}$ is the Lagrangian of the counterterms. In dimensional regularization, the latter  Lagrangian looks like
\be
&& \hspace{-1.0cm} 
 \mathcal{L}_{\rm ct}  
 = \frac{1}{\epsilon} \Big[ - 2 \beta_{\kappa_D^{-2}} R + \beta_{\bar{\lambda}} 
  + \sum_{j=3}^{{\rm N}+2} \sum_{k=3}^{j} \sum_i \beta_{c_{k,i}^{(j)}} \left( \nabla^{2(j-k)} {\cal R}^k \right)_i  
 + \sum_{n=0}^{\mathrm{N} }  \Big( \beta_{a_n} R  \, \Box^n R 
  +
  \beta_{b_n}   R_{\mu\nu}  \Box^n  R^{\mu\nu}   \Big) \Big] ,  
\label{Abeta2}
\ee
where 
$\beta_{\kappa_D^{-2}}, \beta_{\bar{\lambda}}, \beta_{a_n}, \beta_{b_n}, \beta_{c_{k,i}^{(3)}}, \ldots, 
\beta_{c_{k,i}^{(\mathrm{N}+2)}}$
are the beta functions of the theory. 
 Since the one-loop Green functions obtained from the 
effective action must be finite 
when $\epsilon \rightarrow 0$,  
the counterterms Lagrangian is related to the divergent part of the effective Lagrangian by 
$\mathcal{L}_{\rm ct} = -  \mathcal{L}_{\rm div}$. 
The effective action and the beta functions can be calculated using the techniques developed by 
Barvinsky and Vilkovisky in \cite{GBV}. 
Comparing (\ref{actionRen2}) and (\ref{Abeta2}), we find 
\be
 (Z_{\alpha_i} -1) \alpha_i = \frac{1}{\epsilon} \beta_{\alpha_i} \,\,\, \Longrightarrow \,\,\, 
 Z_{\alpha_i} = 1 + \frac{1}{\epsilon} \beta_{\alpha_i} \frac{1}{\alpha_i} \, ,
\ee
where $\alpha_i$ is any of the coupling constants (\ref{coupling}).  
The bare $\alpha_i^{B}$ and the renormalized $\alpha_i$ coupling constants come together in  
$\alpha_i^{B} = \alpha_i \, Z_{\alpha_i}$.
All the $\beta_i$ functions flow to constants in the UV regime, because no divergences come from the vertices in set 1 (\ref{Vertex1}). The resulting beta functions are independent of the coupling constants $\alpha_i$ and they only depend on 
the parameters $s_{k,i}$, $d_{k,i}^{(j)}$ together with the coefficients $a_i$ in the polynomial (\ref{poly}). 
Therefore, it is very simple to solve exactly the renormalization group equations
in the UV regime:
\be
\frac{d \alpha_i}{d t} = \beta_i(\alpha_i) \, , \,\,\,\,\,  t := \log \left( \frac{\mu}{\mu_0} \right). 
\ee
In this way we obtain the following  running for the coupling constants $\alpha_i$:
\be
\alpha_i(\mu) \sim \alpha_i(\mu_0) + \beta_i \, t \, .
\ee
This means, that expressed in the inverse couplings our theory is {\em asymptotically free} (running couplings reach zero in infinite energy scale limit.)
The answer to the question, whether $\beta_{\kappa_D^{-2}}$ and $\beta_{\bar{\lambda}}$ are both positive will be published in a separate paper.

Finally the renormalized one loop effective action in the UV including the finite logarithmic contributions 
and assuming renormalization group initial conditions (\ref{CIde}) reads 
\be
&& \hspace{-1.2cm}
\mathcal{L}_{{\rm 1-loop} }^{\rm Ren} \equiv 
- \frac{2 }{\kappa_D^{2}(\mu_0) } \left( R +
G_{\mu \nu}    \frac{ e^{H(-\Box_{\Lambda})} -1}{ \Box}
 R^{\mu \nu}  \right)
 -  \beta_{\kappa_D^{-2}} \log (\mu^2/\mu_0^2) R 
 -  \beta_{\kappa_D^{-2}} \log (- \Box/\mu^2) R 
 \nonumber  \\
&& \hspace{-1.5cm}
+\frac{1}{2} \sum_{n=0}^{\mathrm{N}}  
 \Big[ \beta_{a_n} \log \left( \frac{\mu^2}{\mu_0^2} \right) R \, \Box^n \, R
+  \beta_{a_n}  R \, \Box^n \, \log \left( \frac{-\Box}{\mu^2} \right)\, R \nonumber \\
&&\hspace{-1.5cm}
   + \beta_{b_n}  \log \left( \frac{\mu^2}{\mu_0^2} \right) R_{\mu \nu} \, \Box^n \, R^{\mu \nu} 
   + \beta_{b_n}  R_{\mu \nu} \, \Box^n \,  \log \left( \frac{ -\Box}{\mu^2} \right)R^{\mu \nu} 
   \Big]\nonumber \\
   && \hspace{-1.5cm}
   +  \frac{1}{2}  \sum_{j=3}^{{\rm N}+2} \sum_{k=3}^{j} \sum_i \left[ c_{k,i}^{(j)} (\mu_0)   
   + \beta_{c_{k,i}^{(j)}} \log \left( \frac{\mu^2}{\mu_0^2} \right) 
   + \beta_{c_{k,i}^{(j)}} \log \left( \frac{-\Box}{\mu_0^2} \right) \right] \left( \nabla^{2(j-k)} {\cal R}^k \right)_i   
    - 2 \tilde{\kappa}_D^{-2}( V_{><} + V_{K})  + \mathcal{L}_{\rm ct} \, . 
   \ee
The renormalization group invariance enables us to simplify the action to the following form
\be
&& \hspace{-1cm}
\mathcal{L}_{{\rm 1-loop} }^{\rm Ren} \equiv 
- \frac{2 }{\kappa_D^{2}(\mu_0) } \left( R +
G_{\mu \nu}    \frac{ e^{H(-\Box_{\Lambda})} -1}{ \Box}
 R^{\mu \nu}  \right)
 \nonumber  \\
&& 
  \hspace{0.6cm}
+\frac{1}{2} \sum_{n=0}^{\mathrm{N}}  
  \Big[   \beta_{a_n}  R \, \Box^n \, \log \left( \frac{-\Box}{\mu_0^2} \right)\, R 
   + \beta_{b_n}  R_{\mu \nu} \, \Box^n \,  \log \left( \frac{ -\Box}{\mu_0^2} \right)R^{\mu \nu} 
   \Big] \nonumber \\
   && \hspace{0.6cm}
    + \frac{1}{2}
    \sum_{j=3}^{{\rm N}+2} \sum_{k=3}^{j} \sum_i \beta_{c_{k,i}^{(j)}} \log \left( \frac{-\Box}{\mu_0^2} \right) \left( \nabla^{2(j-k)} {\cal R}^k \right)_i
  - 2 \tilde{\kappa}_D^{-2}( V_{><} + V_{K})+ \mathcal{L}_{\rm ct} \, , 
\ee
where we assumed, that ${c_{k,i}^{( 3 )} }(\mu_0)= \dots ={c_{k,i}^{( {\rm N} + 2  )} }(\mu_0) =0$.
We can equivalently move such initial conditions in the logarithms of operators from the last line by using the
property $a = \exp ( \log a)$. In two formulas above we used a schematic notation for higher than quadratic in curvature terms, where the action of $\log \left( \frac{-\Box}{\mu_0^2} \right)$ operator should not be understood as a total derivative.

\section{Quantum modern gravity}
In the previous section we showed unitarity around flat spacetime and power-counting 
convergence of the amplitudes beyond one loop.
In this section we quantize the four-dimensional theory defined by (\ref{Action2})
in the path-integral formulation. Using the background field method
we extract the divergent contribution to the one-loop effective action. 
Finally we will show, that the theory doesn't contain any perturbative divergences even at one loop by proper choice of the curvature potential $V_K$. 
For this task, the property (\ref{Vlimit1}) allows us to focus just on the UV  limit of (\ref{Action2}). 
Moreover, without loss of generality, we can consistently fix  
$V_< = 0$, and set to zero the coefficients for the operators 
$a_n  R \, \Box^n R$ and
$b_n  R_{\mu \nu} \Box^n R^{\mu \nu}$ in $\mathcal{L}_Q$ 
(these operators are renormalized only 
if we have one loop divergences.) 
We also assume that $V_{><} =0$,  
because this term is not generated at the quantum level.  
The action of the theory, which we are going to quantize finally reads: 
\be
 \hspace{-0.2cm}
 \mathcal{L}_{\rm g} = -\frac{ 2 }{\kappa^{2}_D}  \sqrt{|g|} \left(R+  G_{\mu \nu}  \frac{ e^{\frac{\gamma_E}{2}} \, p(-\Box_{\Lambda}) }{\Box} R^{\mu \nu} + V_{K} \right) \quad {\rm with} \quad p(z)=z^{\gamma+{\rm N}+1}\,.
     \label{ActionUV}
\ee

In the background field method the metric $g_{\mu\nu}$ is split into a background metric $\bar{g}_{\mu\nu}$ 
and a quantum fluctuation $h_{\mu\nu}$
\be
g_{\mu\nu} = \bar{g}_{\mu\nu} + h_{\mu \nu}. 
\label{BFM}
\ee
Sometimes below we will denote these metrics by $g$, $\bar{g}$ and $h$ without writing covariant indices explicitly. Additionally from now on we will not speak about the full metric $g$ and for simplicity of notation the background metric will be denoted by $g$, hoping that this will not lead to any confusion.
In our theory diffeomorphism gauge invariance is present and this is the reason, why we have to fix the gauge and in the quantization procedure we introduce FP ghosts.
The gauge fixing and FP-ghost actions are as follows  
\be
&& \hspace{-1.2cm} 
S_{\rm gf}  = \int \! d^D x \sqrt{ - {g} } \, \chi_{\mu} \, C^{\mu \nu} \, \chi_{\nu} \, , 
\,\,\, 
\chi_{\mu} = {\nabla}_{\sigma} h^{\sigma}_{\mu} - \beta_g  {\nabla}_{\mu} h \, , 
\,\,\,\,   
C^{\mu \nu} \! = - \frac{1}{\alpha_g}
 \left( {g}^{\mu \nu} \Box 
 + \gamma_g {\nabla}^{\mu} {\nabla}^{\nu} - {\nabla}^{\nu} {\nabla}^{\mu} \right)  
 \Box_{\Lambda}^{  {\rm N} + \gamma}  , \nonumber \\
&& \hspace{-1.2cm}  
S_{\rm gh} = \int \! d^D x \sqrt{ - {g} } \left[  \bar{C}_{\alpha} \, M^{\alpha}{}_{\beta} \, C^{\beta} 
+ b_{\alpha} C^{\alpha \beta} b_{\beta} \right] , 
\,\,\,\, 
M^{\alpha}{}_{\beta} = \Box \delta^{\alpha}_{\beta}  
+{\nabla}_{\beta} {\nabla}^{\alpha}  - 2 \beta_g {\nabla}^{\alpha} {\nabla}_{\beta} . 
\label{shapiro3a}
\ee
In (\ref{shapiro3a}) 
we used a covariant  
gauge fixing with weight function $C^{\mu\nu}$  \cite{shapiro}. 
The gauge fixing parameters
$\beta_g$ and $\gamma_g$ are dimensionless,  while $[\alpha_g] = M^{4-D}$. 
We notice right here that in our theory the beta functions are independent of these gauge parameters   
(see \cite{shapiro} for a rigorous proof.)

The partition function with the right functional measure 
compatible with BRST invariance \cite{Anselmi:1991wb, Anselmi:1992hv, Anselmi:1993cu} reads 
\be
&& \hspace{-0.5cm}
Z[g] = \int \!  
\mu(g, h )
\prod_{\mu \leqslant \nu} \mathcal{D}h_{\mu\nu} \prod_{\alpha} \mathcal{D} \bar{C}_{\alpha}
 \prod_{\beta}  \mathcal{D}{C}^{\beta}
 \prod_{\gamma}  \mathcal{D}{b}_{\gamma} \, 
 e^{i \int d^Dx \left[  \mathcal{L}_{\rm g} +  \mathcal{L}_{\rm gf} + \mathcal{L}_{\rm gh}   \right]} \,  .
\label{QG}
 \ee
 At one loop we can evaluate the functional integral and express the partition function as a product of determinants,
 namely
  \be
  Z[g] = e^{i S_{\rm g}[{g}]}   \left\{   \left. {\rm Det}\left[ 
   \frac{\delta^2 (S_{\rm g}[g+h] + S_{\rm gf}[g+h]) }{\delta h_{\mu \nu} \delta h_{\rho \sigma}} \right\vert_{h=0} \right] \right\}^{-\frac{1}{2}} 
 ({\rm Det} \, M^{\alpha}{}_{\beta} ) \, \, 
({\rm Det} \, C^{\mu \nu} )^{\frac{1}{2}}  \, . 
 \nonumber 
 \ee
 By symbol $S_{\rm g}[g]$ we understand classical functional of the gravitational action of the theory.
To calculate the one loop effective action we need first to expand the action plus the gauge-fixing term to the second order in the quantum fluctuation $h_{\mu \nu}$
\be
 \hat{H}^{\mu \nu , \rho \sigma}= \frac{\delta^2 S_{\rm g}}{ \delta h_{\mu \nu}\delta h_{\rho \sigma}}\Bigg|_{h=0} 
 \!\!\! 
 +  \frac{\delta \chi_{\delta}}{\delta h_{\mu \nu}} \, C^{\delta \tau } \, 
 \frac{\delta \chi_{\tau}}{\delta h_{\rho \sigma}}\Bigg|_{h=0} 
  . \label{H}
\ee
The explicit calculation of the full operator $ \hat{H}$ goes beyond the scope of this paper, because here 
we are interested only in showing finiteness of the theory (\ref{Action2}).
Therefore, from here on 
besides assuming the polynomial in (\ref{poly}) to be $z^{\gamma +{\rm N}+1}$, we restrict to $D=4$, hence ${\rm N}=0$. In this case, as explicitly showed in the previous section, all the beta functions vanish except for 
\be 
\beta_{R_{\mu\nu}^2} \, \,\, {\rm and} \,\, \,
\beta_{R^2}. 
\label{beta}
\ee

The Lagrangian density (\ref{ActionUV}) for odd values of the integer $\gamma$ (this technical requirement avoids the absolute value in the action defined along the real axis) reduces to 
\be
 && \hspace{-0.7cm}
 \mathcal{L}_{\rm g} = - 2 \kappa^{-2}_4  \sqrt{|g|} \left( R + G_{\mu \nu}   
 \, \frac{e^{\frac{\gamma_E}{2}} \, \Box_{\Lambda}^{\gamma}}{\Lambda^2} \, R^{\mu \nu} + V_{K} \right)\nonumber \\
 && \hspace{-0.2cm}
 =   - 2 \kappa^{-2}_4 \sqrt{|g|} \Big[ 
  R + \omega_1 \, R \, \Box^{\gamma} \,  R +
   \omega_2\, R_{\mu \nu} \, \Box^{\gamma} \,  R^{\mu \nu}  
+  s_1 \, R^2 \, \Box^{\gamma-2 } \,  R^2 +
  s_2 \, R_{\mu\nu} R^{\mu\nu} \, \Box^{\gamma-2 } \,  R_{\rho \sigma} R^{\rho \sigma}  \Big]  ,
  \label{simply} \\
  &&  \hspace{-0.7cm}
  {\rm where}\quad\omega_2  = - 2 \, \omega_1  = e^{\gamma_E/2}/\Lambda^{2 \gamma+2}\,. \label{omega12relation}
  \ee
  Even with these simplifications the operator $\hat{H}$ is very complicated, but for a rigorous proof of  finiteness 
it is sufficient to calculate the second variation of the terms in the potential $V_K$. In the second line above we already listed the minimal set of operators (in $D=4$) needed to make the theory finite. We will show by an explicit computation that the tensorial structure of these terms is proper for our goal and we will find values for the coefficients $s_1$ and $s_2$.
Since we are interested in the contributions to the one loop beta functions, then
the two local operators coming from the potential (second line in (\ref{simply})) can only
give quadratic contributions (in gravitational curvatures) to the variation and do not interfere with the other operators (to this order in curvature expansion).
In other words only the two Feynman diagrams linear in $s_1$
and  $s_2$
give a non-vanishing contribution to the beta functions (\ref{beta}). 
Since we are interested in the finite theory of quantum gravity, then we can concentrate on
these terms.

Following \cite{shapiro} we can recast (\ref{H}) in the following compact form
\be
&& \hspace{-1.2cm}
\hat{H}^{\mu \nu, \alpha \beta} = \left(  \frac{\omega_2}{4} g^{\mu (\rho} g^{\nu) \sigma}
- \frac{\omega_2 (\omega_2+ 4 \omega_1)}{16 \omega_1} g^{\mu \nu} g^{\rho \sigma} 
\right) \times 
 \label{acca}
\Big\{  
\delta^{\alpha\beta}_{\rho\sigma} \Box^{\gamma+2} 
+ V_{\rho\sigma}{}^{\alpha\beta,\lambda_1  \dots \lambda_{2 \gamma +2}} 
\nabla_{\lambda_1}  \dots \nabla_{\lambda_{2 \gamma+2 }} +
\\
&& \hspace{0.35cm}
+ W_{\rho\sigma}{}^{\alpha\beta,\lambda_1  \dots \lambda_{2 \gamma + 1}} 
\nabla_{\lambda_1} \dots \nabla_{\lambda_{2 \gamma+1}}  
+ U_{\rho\sigma}{}^{\alpha\beta,\lambda_1 \dots \lambda_{2 \gamma }} 
\nabla_{\lambda_1}  \dots \nabla_{\lambda_{2 \gamma}} +O(\nabla^{2\gamma-1})  \Big\}     \, , 
\nonumber 
\ee
where $\delta_{\mu\nu}^{\rho \sigma} \equiv \delta_\mu^{(\rho}\delta_\nu^{\sigma)} \equiv \frac{1}{2} \left(\delta_\mu^{\rho}\delta_\nu^{\sigma}+\delta_\mu^{\sigma}\delta_\nu^{\rho} \right)$ 
and  the tensors $V, W$ and $U$ depend on 
curvature tensors of the background metric and its covariant derivatives.
In (\ref{acca}) the pre-factor in round brackets (called de Witt metric) does not give any contribution to the divergences and therefore it
can be omitted. The tensor $V$ is linear in curvature tensor, while the 
tensor $U$ takes contributions quadratic in curvature ($R^2$). We  obtain expressions for $U,\,V$ and $W$ tensors by contracting with the inverse de Witt metric and extracting at the end covariant derivatives. They have the canonical position of matrix indices (two down followed by two up) thanks to the application of this metric in the field fluctuation space.
We will concentrate mostly on the tensor $U$, because only that one carries in contributions to the divergent part from the potential in curvature in our case.
Corresponding formulas for the tensor before the multiplication by the inverse de Witt metric will be decorated with a prime after the name of this tensor. 

Here the goal is to make the theory finite engaging a sufficient number of hit men to kill the one-loop contributions to the beta functions. 
 As explained above, the two operators in the local potential (\ref{simply}) 
\be
 s_1  R^2  \Box^{\gamma-2 }   R^2\, \,\,\,  {\rm and} \,\,\, \,
s_2 R_{\mu\nu} R^{\mu\nu}  \Box^{\gamma-2 }  R_{\rho \sigma} R^{\rho \sigma} \, , 
\label{killers2}
\ee 
can be good murderers of the beta 
functions for the two terms, that are quadratic in curvatures. The reader can easily see, that 
they still do their job despite their quite simple structure. 
Importantly they really can kill both beta functions from (\ref{beta}), because  
their contributions do not vanish and have proper tensorial structure in curvature tensors.
In the last part of the paper we will give the details of this pretty short computation.

The one-loop effective action is 
defined by 
\cite{shapiro}
\be
&& \hspace{-0.4cm}
\Gamma^{(1)}[g] = - i  \log Z[g] = 
\label{Gamma1}
 S_{\rm g}[g] 
+ \frac{i}{2}   \ln {\rm Det}
(\hat{H}) - i  \ln {\rm Det}(\hat{M})
 - \frac{i}{2} \ln
{\rm Det}(\hat{C})  . \nonumber 
\ee
Once the relevant contributions to the operator $\hat{H}$ are known we can apply 
the Barvinsky-Vilkovisky method
\cite{GBV} to extract the divergent part of $  \ln
{\rm Det} (\hat{H}^{\mu\nu,\alpha\beta})$. 

Now using the identity
$\ln {\rm Det} (\hat{H}) = {\rm Tr} \ln \hat{H}$ we have the contribution from killers to the one-loop action
\be
&& \hspace{-0.4cm}
{\rm Tr} \ln \hat{H}^{\mu \nu, \alpha \beta} \supset 
  {\rm Tr} \left( 
U_{\rho \sigma}{}^{ \alpha \beta,\lambda_1 \dots \lambda_{2 \gamma }} 
\nabla_{\lambda_1}  \dots \nabla_{\lambda_{2 \gamma}}     
\frac{1}{\Box^{\gamma+2}} \right) 
+ O(\nabla^k {\rm Riem}^l,k+2l>4)\, .
\label{tracciaT}
\ee
We are interested in finding a finite theory of quantum gravity, therefore we  concentrate on the
$U$ tensor, which contains operators quadratic in the curvature, but linear in $s_1$ and $s_2$. 
The resulting beta functions are linear in the parameters 
$s_1$ and $s_2$, because from the killer operators (\ref{killers2}) we do not get divergent 
one-loop Feynman graphs, if we have more than one external leg. Namely we write schematically that
\be
&&  \hspace{-0.35cm}
\beta_{R^2} := a_{1} s_1 + a_2 s_2  + c_1 \, , \nonumber  \\
&& \hspace{-0.35cm}
 \beta_{R_{\mu\nu}^2} := 
  b_2 s_2 + c_2 \, .  
\label{beta2}
\ee
We will see by explicit calculation, that the operator with scalar curvatures does not give rise to contribution to the second beta function as reported above.
The coefficients $a_1$, $a_2$, $b_2$  
come from traces in (\ref{tracciaT}).
What we need to show finiteness of the theory is to find the trace of the operators in (\ref{killers2}), when included to $\hat{H}$. Traces of all the other terms present in the second variational operator $\hat{H}$
only give contribution to the constants $c_1$ and $c_2$. Due to dimensional reasons the coefficients $a_1$, $a_2$, $b_2$ are functions of kinetic part parameters $\omega_1$ 
and $\omega_2$.
The two quartic operators in (\ref{killers2}) are independent, they will give a different non zero contribution 
to the beta functions. The constants $c_1$ and $c_2$ come from the contributions of other vertices and propagators in (\ref{simply}). More generally they can be viewed as functions 
of a dimensionless ratio $\omega_2/\omega_1$, which is equal $-2$ in our theory. 
Looking back at (\ref{beta2}) we immediately conclude, that the beta functions can be made vanish for real values of the parameters $s_1$ and $s_2$ such that 
\be 
s_1 = -\frac{c_1 b_2 - c_2 a_2 }{a_1b_2} \,\,\,\,\, \mbox{and} \,\,\,\, 
s_2 = -\frac{c_2}{b_2} \, . 
\ee

\subsection*{Explicit computation of the coefficients $s_1$ and $s_2$} 

In this subsection we will explicitly derive the coefficients $a_1$, $a_2$ and $b_2$. However we will not need to find functions $c_1$ and $c_2$ to show the finiteness of the theory. 
For this task we first need the second variation of the operators (\ref{killers2}) and then after contraction 
with the inverse de Witt metric we will be ready to evaluate the traces in (\ref{tracciaT}) using the Barvinsky-Vilkovisky technology. We emphasize moreover, that 
all the couplings involved in the expression for 
the beta functions do not run with the energy scale as it was explained in the previous section.

Let us now start computing explicitly the second variation of the first operator quartic in the curvature:
$R^2 \Box^{\gamma-2} R^2$.
For our purposes we need $2\gamma$ covariant derivatives acting between metric fluctuations,
while the outcome of the variation must
contain terms quadratic in the background curvature. For this operator 
the computation leading to contributions to $U'$ 
is exactly the same like for the case of the higher derivative kinetic term 
with $\square^\gamma$ ($U$ in (\ref{acca}) is obtained by multiplication of $U'$ with the inverse 
de Witt metric). 
Here however the result is additionally multiplied by $R^2$. 
Luckily we have no problem with self-adjointness, Leibniz expansion, 
neither with commutations of derivatives for this contribution. Exploiting integration by parts under the integral, the final expression for the second variation is  made of 
just four terms, namely 
\be
&&\hspace{-1cm} 
 \delta^2 \! \left(R^2 \square^{\gamma-2} R^2 \right) 
 = 8 R^2 \left( h \square^\gamma h - h \nabla^\mu \nabla^\nu \square^{\gamma-1} h_{\mu\nu} 
-  h_{\mu\nu} \nabla^\mu \nabla^\nu \square^{\gamma-1} h + h_{\mu\nu} \nabla^\mu \nabla^\nu \nabla^\rho \nabla^\sigma \square^{\gamma-2} h_{\rho\sigma} \right) 
\nonumber  \\
&& \hspace{1.75cm} 
+ O \left( \nabla^k \mathrm{Riem}^l, k + 2l > 4 \right)  . 
\ee
Thus the operator of the second variational derivative reads as follows 
\be
&& 
\hat{H}^{\alpha\beta,\,\zeta\delta}
 = 8 R^2 \left( g^{\alpha\beta} g^{\zeta\delta} \square^\gamma - g^{\alpha\beta} \nabla^\zeta \nabla^\delta \square^{\gamma-1}  
 - g^{\zeta\delta} \nabla^\alpha \nabla^\beta \square^{\gamma-1}  + \nabla^\alpha \nabla^\beta \nabla^\zeta \nabla^\delta \square^{\gamma-2} \right)
  \nonumber  \\
  && \hspace{1.5cm}
+ O \left( \nabla^k \mathrm{Riem}^l, k + 2l > 4 \right)  . 
\ee
In general situation the contribution of $U'$ to the operator $\hat{H}$ in (\ref{acca}) is
with four derivatives: 
\be
\hspace{-0.3cm}
U'^{\,\alpha\beta,\,\zeta\delta,\,\lambda_1\lambda_2\lambda_3\lambda_4} \nabla_{\lambda_1} \nabla_{\lambda_2} \nabla_{\lambda_3} \nabla_{\lambda_4} \square^{\gamma-2} \subset \hat{H}^{\alpha\beta,\,\zeta\delta}.
\ee
In our special case we find, that 
$U'$ (which comes with $\square^{\gamma-2}$) is: 
\be
&& \hspace{-1.2cm}
U'^{\,\alpha\beta,\,\zeta\delta,\,\lambda_1\lambda_2\lambda_3\lambda_4} 
 = 8 R^2 \left( g^{\alpha\beta} g^{\zeta\delta} g^{\lambda_1\lambda_2} g^{\lambda_3\lambda_4} \right. 
- g^{\alpha\beta} g^{\lambda_3\lambda_4} g^{\zeta\lambda_1} g^{\delta\lambda_2} - g^{\zeta\delta} g^{\lambda_3\lambda_4} g^{\alpha\lambda_1} g^{\beta\lambda_2} 
  \left. + g^{\alpha\lambda_1} g^{\beta\lambda_2} g^{\zeta\lambda_3} g^{\delta\lambda_4} \right) 
  \nonumber \\
  && \hspace{1.7cm}
  + O\left( \nabla^k \mathrm{Riem}^l, k + 2l > 4 \right) \, .
\label{Ufirstkiller}
\ee
The inverse de Witt metric in our theory has the following compact form
\be
C_{\eta\theta,\,\alpha\beta} = y_1 \, g_{\eta\theta} g_{\alpha\beta} + y_2 \, g_{\eta(\alpha} g_{\beta)\theta} \,  , \,\,\,\, y_1=-\frac{x_1}{x_2(Dx_1+x_2)} \, ,  \quad \mbox{and} \quad y_2=\frac{1}{x_2},  
\ee
where $x_1$ and $x_2$ are the corresponding coefficients of the de Witt metric as appearing in (\ref{acca}). Hence in $D=4$ we have explicitly
\be
&& y_1 = -\frac{4\omega_1+\omega_2}{2\omega_2(3\omega_1+\omega_2)} \, ,
\nonumber \\  
&& y_2 = \frac{2}{\omega_2}\,.
\label{invdewitt}
\ee
Multiplying (\ref{Ufirstkiller}) from the left by the inverse metric in field space we finally get the following contribution to $U$,  
\be
&& \hspace{-1.4cm}
U_{\eta\theta}{}^{\zeta\delta,\,\lambda_1\lambda_2\lambda_3\lambda_4}  
 = 4 R^2 \left\{ y_2 \, 
g^{\lambda_1} {}_\eta \, g^{\lambda_2} {}_\theta \, (g^{\zeta\lambda_3} g^{\delta\lambda_4} - g^{\zeta\delta} g^{\lambda_3\lambda_4})  
\right. 
 - g_{\eta\theta} \left[ y_1 \, g^{\zeta\lambda_3} g^{\delta\lambda_4} g^{\lambda_1\lambda_2} - \left((4 \,y_1 + y_2) g^{\zeta\lambda_1} g^{\delta\lambda_2}  \right. \right. \nonumber \\
&& \hspace{1.1cm}
\left.\left.\left. -
(3 y_1 + y_2) g^{\zeta\delta} g^{\lambda_1\lambda_2}\right) g^{\lambda_3\lambda_4} \right] \right\} + O\left( \nabla^k \mathrm{Riem}^l, k + 2l > 4 \right) . 
\ee

We are sure that at this order in curvature there is no interference  between $U$ and the matrix operators  $V$ and $W$, because the contribution above 
is already quadratic in curvature.

Let us calculate the variation of the second killer in (\ref{killers2}).  
The computation of this variation is only a bit more involved, 
because of the tensorial structure. 
The main task of the second murderer is to kill the beta function for the tensorial operator
$\beta_{R_{\mu \nu}^2}$. Its structure was engineered specially for this.
There is of course a possibility that this term also contributes
 partially to the beta function for the scalar curvature squared term. But this 
only means, that our chosen two murderers must create a linear combination to achieve their 
goals. The contribution of the second killer to the $U'$ operator is made of nine terms. 
Here we have  the linear superposition principle at work, i.e. the contribution from the sum of 
terms is the sum of each term contributions. 
The form of the second variation for this operator is given explicitly by: 
\be
&& \hspace{-1cm} \delta^2 \! \left(R_{\mu\nu}^2 \square^{\gamma-2} R_{\mu\nu}^2 \right) 
 = 2  R^{\mu\nu} R^{\rho\sigma} \left( h_{\mu\nu} \square^\gamma h_{\rho\sigma} +
  \right. 
h_{\mu\nu} \nabla_\rho \nabla_\sigma \square^{\gamma-1} h 
+ h \nabla_\rho \nabla_\sigma \square^{\gamma-1} h_{\mu\nu} 
 - 2 h_{\mu\nu} \nabla_\rho \nabla^\tau \square^{\gamma-1} h_{\sigma\tau} 
\nonumber \\
&& \hspace{-1cm} 
-2 h_{\sigma\tau} \nabla_\rho \nabla^\tau \square^{\gamma-1} h_{\mu\nu} +
h \nabla_\mu \nabla_\nu \nabla_\rho \nabla_\sigma \square^{\gamma-2} h 
- 2 h \nabla_\mu \nabla_\nu \nabla_\rho \nabla^\tau \square^{\gamma-2} h_{\sigma\tau} 
- 2  
h_{\sigma\tau} \nabla_\mu \nabla_\nu \nabla^\rho 
\nabla^\tau \square^{\gamma-2} h \nonumber \\
&& \hspace{-1cm} 
+ 4  
h_{\mu\tau} \nabla_\nu \nabla_\sigma \nabla^\tau \nabla^\upsilon \square^{\gamma-2} h_{\rho\upsilon} )
+ O\left( \nabla^k \mathrm{Riem}^l, k + 2 l > 4 \right) . 
\ee
Thus the contribution to the operator of the second variational derivative derived from this term is:
\be
&& \hspace{-1cm}
\hat H^{\alpha\beta,\,\zeta\delta}
= 2 \Big( R^{\alpha\beta} R^{\zeta\delta} \square^\gamma 
+ R^{\alpha\beta} R^{\rho\sigma} g^{\zeta\delta} \nabla_\rho \nabla_\sigma \square^{\gamma-1} 
+ R^{\zeta\delta} R^{\rho\sigma} g^{\alpha\beta} \nabla_\rho \nabla_\sigma \square^{\gamma-1}  
- 2 R^{\alpha\beta} R^{\rho\zeta} \nabla_\rho \nabla^\delta \square^{\gamma-1} 
\nonumber \\
&& \hspace{-1cm} 
- 2 R^{\zeta\delta} R^{\rho\alpha} \nabla_\rho \nabla^\beta \square^{\gamma-1}  
+ R^{\mu\nu} R^{\rho\sigma} g^{\alpha\beta} g^{\zeta\delta} \nabla_\mu \nabla_\nu \nabla_\rho \nabla_\sigma \square^{\gamma-2}
 - 2 R^{\mu\nu} R^{\rho\zeta} g^{\alpha\beta} \nabla_\mu \nabla_\nu \nabla^\rho \nabla^\delta \square^{\gamma-2}  \nonumber \\
&& \hspace{-1cm} 
- 2 R^{\mu\nu} R^{\rho\alpha} g^{\zeta\delta} \nabla_\mu \nabla_\nu \nabla^\rho \nabla^\beta \square^{\gamma-2} 
+ 4 R^{\alpha\nu} R^{\zeta\sigma} \nabla_\nu \nabla_\sigma \nabla^\beta \nabla^\delta \square^{\gamma-2} \Big)
+ O\left( \nabla^k \mathrm{Riem}^l, k + 2 l > 4 \right) , 
\ee
and the contribution to the $U'$ operator is following:
\be
&& \hspace{-0.5cm} U'^{\,\alpha\beta,\,\zeta\delta,\,\lambda_1\lambda_2\lambda_3\lambda_4}
= 2 \Big( R^{\alpha\beta} R^{\zeta\delta} g^{\lambda_1\lambda_2} g^{\lambda_3\lambda_4} 
+ R^{\alpha\beta} R^{\lambda_1\lambda_2} g^{\zeta\delta} g^{\lambda_3\lambda_4} + R^{\zeta\delta} R^{\lambda_1\lambda_2} g^{\alpha\beta} g^{\lambda_3\lambda_4}  
\nonumber \\
&& \hspace{-0.5cm} 
- 2 R^{\alpha\beta} R^{\lambda_1\zeta} g^{\lambda_3\lambda_4} g^{\delta\lambda_2} - 2 R^{\zeta\delta} R^{\lambda_1\alpha} g^{\beta\lambda_2} g^{\lambda_3\lambda_4} 
+ R^{\lambda_1\lambda_2} R^{\lambda_3\lambda_4} g^{\alpha\beta} g^{\zeta\delta} - 2 R^{\lambda_1\lambda_2} R^{\lambda_3\zeta} g^{\alpha\beta} g^{\delta\lambda_4}  \nonumber \\
&& \hspace{-0.5cm} - 2 R^{\lambda_1\lambda_2} R^{\lambda_3\alpha} g^{\zeta\delta} g^{\beta\lambda_4} +  4 R^{\alpha\lambda_1} R^{\zeta\lambda_2} g^{\beta\lambda_3} g^{\delta\lambda_4}  
\Big)
+ O \left( \nabla^k \mathrm{Riem}^l, k + 2 l > 4 \right) .
\ee

We decided not to write the full expression for $U$ operator in this case, because of its length. The structure is very similar to the previously encountered one for the first operator with scalar curvatures.

Of course the contribution of the operators (\ref{killers2}) to the divergent 
part of the effective action is quadratic in curvature and linear in the 
coefficients $s_1$ and $s_2$. 
In consequence we restricted the computation of the second variation to order quadratic in curvature, because by performing the functional traces and taking their divergent part, we only increase or remain with the same power  in curvatures.
These contributions to the divergent part depend also on the coefficients $\omega_1$ and $\omega_2$ 
multiplying the operators $R\, \square^\gamma R$ and $R_{\mu\nu} \, \square^\gamma R^{\mu\nu}$
in the action. 
The dependence on $\omega_1$ and $\omega_2$ appears here, 
when we multiply by the inverse deWitt metric.  
The parameters $\omega_1$ and $\omega_2$ enter in the denominators of the coefficients $y_1$ and $y_2$ as shown in (\ref{invdewitt}). 
However, this additional nonlinear dependence is not a problem for us, because we only want to determine  
the coefficients $s_1$ and $s_2$ in front of the killers. And the killers give to the beta functions only a linear contribution in $s_1, s_2$, 
so at the end we only have to solve a two-dimensional system of linear equations (\ref{beta2}).

Although the finiteness of the theory as presented here is based on a particular 
choice of the coefficients $s_i$ ($i=1,2$), our result has a universal character.
We can add more operators to the action (\ref{Action2}), therefore increasing the dimension of the parameter space, likely maintaining the theory finite.  They will all be added to the curvature potential $V_K$.
In this paper we mainly consider the minimal curvature potential, able to kill the beta functions for 
the counterterms of operators quadratic in curvatures
$R_{\mu\nu}^2$ and $R^2$. However in general we can add a maximal, but finite number of operators  
still having a finite quantum gravity. This point is expanded towards the end of this section.

Now we want to report final results about traces of the killers.  
All contributions are usually multiplied by the divergent coefficient depending on the 
regularization scheme (in report paper \cite{GBV} this is $i \log(L^2)/(16 \pi^2)$, 
where $L$ is a cutoff scale). As we expected the first killer contributes only to the divergent term proportional to the operator 
$R^2$.
Tracing the first killer we find the additional non zero factor
\be
\frac{12}{3\omega_1 + \omega_2} \, , 
\ee
originating basically from the deWitt metric.

The trace of the second killer has a bit more interesting structure and 
it is correct to expect, that it gives contributions to both beta functions in (\ref{beta}). 
For the divergent contribution proportional to $R^2$ the fraction coming from tracing is: 
\be
 \frac{-10 \omega_1 + \omega_2}{6 \omega_2 (3 \omega_1 + \omega_2)}  \, . 
 \ee
More importantly there is a nonzero factor that multiplies
the divergent part with a tensorial operator $R_{\mu\nu}R^{\mu\nu}$, namely  
\be
\frac{20 \omega_1 + 7 \omega_2}{3\omega_2(3\omega_1 + \omega_2)} \, . 
\ee
We see, that in general all these three contributions are non zero. It is significant to observe, that the condition to have a unitary 
theory (\ref{omega12relation}) does not deny conditions for these terms to be non-vanishing. Hence it is possible to have a unitary, super-renormalizable and finite theory.
 
In summary the 
functional traces for the special operators (\ref{killers2}) added to our action amount to
\be
\mathrm{Tr} \log \hat{H}_{K1} = i \frac{\log(L^2)}{16 \pi^2} \frac{12 R^2}{3\omega_1 + \omega_2}
\ee
from the first killer and
\be
\mathrm{Tr} \log \hat{H}_{K2} 
= i \frac{\log(L^2)}{16 \pi^2} \left( \frac{-10 \omega_1 + \omega_2}{6 \omega_2 (3 \omega_1 + \omega_2)} R^2 + \frac{20 \omega_1 + 7 \omega_2}{3\omega_2(3\omega_1 + \omega_2)} R_{\mu \nu}^2 \right) \nonumber 
\ee
from the second killer.

Next we use the identification  $\frac{\log(L^2)}{16 \pi^2}=-\frac{1}{8\pi^2} \frac{1}{\epsilon}$ (from formula (4.38) in \cite{GBV}) to find finally, that the coefficients $a_1$, $a_2$ and $b_2$ are:
\be
a_1=-\frac{1}{8\pi^2}\frac{6}{3\omega_1 + \omega_2},\quad a_2=-\frac{1}{8\pi^2}\frac{-10 \omega_1 + \omega_2}{12 \omega_2 (3 \omega_1 + \omega_2)}\quad{\rm and}\quad b_2=-\frac{1}{8\pi^2}\frac{20 \omega_1 + 7 \omega_2}{6\omega_2(3\omega_1 + \omega_2)}\,.
\ee

The condition $(3 \omega_1 + \omega_2)\neq 0$ is necessary, if we want the same scaling for all the components of the propagator, namely $k^{2\gamma+4}$ UV. This particular combination was for the first time 
pointed out in quadratic gravity by Stelle in 1977 \cite{Stelle}.  
Moreover, we of course require, that $\omega_2 \neq 0$. All these conditions originate from the coefficients of the inverse deWitt metric. They are all satisfied, when the standard conditions for the theory hold true. Therefore in this situation all our results for traces are well defined. This not only works in $D=4$, but is 
consistently generalized to other dimensions with full agreement.

If $\gamma \geqslant 3$, then the divergent contributions come only at one loop. 
As already stressed, we assume the polynomial (\ref{poly})
to be restricted to the first monomial $z^{\gamma +1}$, then only the vertices proportional to 
$\omega_1$ and $\omega_2$ in (\ref{simply}) give a contribution to the divergences.
In particular there  are no divergent contributions to the cosmological constant or the Ricci scalar term. 
Conversely, there maybe present other terms quartic or higher  
in curvature. By adding these other terms we must be careful to do not spoil the condition for the renormalizability of the theory, namely the number of derivatives of metric in these terms must be bounded by $2\gamma+4$.
However, terms higher than quartic do not source the divergent contributions to one-loop effective action in four dimension. It turns out, that
the divergences in dimension four concentrate only in $R^2$ and $R_{\mu\nu}^2$ terms. 
To calculate the contributions of all quartic terms is really a formidable 
task. It is very difficult even to algebraically classify all appearing terms. We assume, that they come with coefficients $(\omega_3, \ldots,  \omega_m)$ ($m$ is some combinatoric function of $\gamma$, but for given $\gamma$ this is always a finite number.)
In full generality we can only say, that from all these terms (with exception of two killers) the contribution to the divergent terms is encoded in two functions $c_1(\omega_1,\omega_2, \ldots ,\omega_m)$ and $c_2(\omega_1,\omega_2, \ldots ,\omega_m)$ depending linearly on the coefficients of the quartic terms and quite non-linearly on 
$\omega_1$ and $\omega_2$. 
And here this is already much more general than modest initial goal to compute only the full dependence on $\omega_1$ and $\omega_2$. 
We know, that because of many theorems (like covariance of counterterms \cite{AnselmiN}) 
for every gravitational theory these two functions must exist (and in principle are computable). 
 
 Now, 
 using our two special killers (\ref{killers2}) from the set of quartic operators, 
 we can make the theory finite for whatever set of coefficients $\omega_1$, $\omega_2$ 
 ($\omega_2 \neq 0$ and $3\omega_1+\omega_2 \neq 0$) and $\omega_3$ to $\omega_m$. 
For every value of them the functions $c_1$ and $c_2$ are computable and take particular values. 
Once their values are known, then the coefficients $s_1$ and $s_2$ are given by the formulas:
\be 
&& 
 s_1 = \frac{ 2\pi^2( 3\omega_1 + \omega_2)(40  c_1  \omega_1 + 10  c_2 \omega_1 + 14  c_1  \omega_2 
 - c_2 \omega_2)}{          3(20\omega_1 + 7\omega_2)} \, ,  \,\,\,\,\,\,   \nonumber \\
&& 
s_2 = \frac{ 48\pi^2 c_2 \omega_2(3\omega_1 + \omega_2)}{20\omega_1 + 7\omega_2} . 
\label{solcoeffs}
\ee 
 The two conditions on the parameters $s_1$ and $s_2$ among $m+2$ parameters of the theory 
 in the quartic in curvature sector make the theory of quantum gravity finite.

\section{Conclusions \& Remarks}  

\subsection*{The Results}
In this paper we advanced the most general gravitational theory compatible with 
super-renormalizability or finiteness together with unitarity. 
The theory is defined by equations (\ref{Action0}), (\ref{operators}) and (\ref{K0}) in a multidimensional spacetime and by (\ref{Action2}) in $D=4$. 
The action consists of a non-polynomial kinetic term with asymptotic polynomial behavior and a local potential of the curvature $O({\cal R}^3)$. 
It has been explicitly shown, that quantum divergences only occur at one loop and the theory is 
super-renormalizable in any dimension. If we make a specific choice for a restricted number of parameters in the potential like in (\ref{solcoeffs}), then all the beta functions can be made to vanish and the theory turns out to be finite. The result has been explicitly proven in dimension four, but can be easily generalized to any dimension.

\subsection*{Four dimensional theory in a nutshell} 
In dimension four the whole situation is simple 
 to describe. 
The highest derivative terms in the kinetic part of
the action come from the form factor and 
are of the type ${\cal R} \, \square^{\gamma}{\cal R}$. 
The two coefficients
in front of them give the shape to the denominators of the beta functions, because these two
terms determine the UV behavior of the propagator. 
For renormalizability $\gamma\geq 0$. If $\gamma=0$, then we have only renormalizability and 
the divergences must be absorbed at every loop order. For $\gamma=1$ we have
3-loop super-renormalizability (so no divergences at $4$ loops). For
$\gamma=2$ we have 2-loop super-renormalizability and finally starting
from $\gamma=3$  
we have one-loop super-renormalizability. Therefore, 
quantum divergences can appear at most at one loop. 

Now increasing
the value of $\gamma$ does not improve the situation, however we can
ask easily for finiteness of the theory. Divergences at one-loop cause
the need for the renormalization of only the following four terms: $\bar{\lambda}$,
$R$, $R^{2}$ and $R_{\mu\nu}^{2}$. 
Let us summarize them in order
saying which operators contribute to which divergences. We will describe
the operators by giving their total number of derivatives acting on
the metric tensor and giving the number of Riemann curvature tensors involved.

About the running of the cosmological constant only the operators quadratic in the curvature have impact. 
The terms with $2\gamma$ derivatives give a contribution linearly proportional to their frontal
coefficients, while terms with $2\gamma+2$ derivatives
give contributions quadratically dependent on their coefficients. 
As it will be shown elsewhere, it is not possible to find such a combination
for non zero values of coefficients of these operators to make the cosmological beta function vanish
\cite{MathStructure}. 

The running of the Planck scale parameter is simpler. There are two
contributions, which are linearly proportional to frontal coefficients of
the corresponding terms. From the quadratic in curvature terms there is
one relevant type of terms with $2\gamma+2$ derivatives. Here also 
the potential contributes with terms cubic in curvature and again with $2\gamma+2$ derivatives. 
Therefore, it is possible to
solve one linear equation expressing the condition for the vanishing of
the beta function and to find the values of the coefficients of cubic terms. 
One cubic killer probably does the job of killing this beta function.

The beta functions for terms quadratic in curvatures are complicated, 
but all of these contributions come from the terms with $2\gamma+4$
derivatives on the metric. First, there are contributions coming from the
highest derivative terms in the kinetic part, so with $2\gamma+4$
derivatives (from the two operators of the type ${\cal R}\square^{\gamma}{\cal R}$.) Dependence
on their coefficients is given by quite nonlinear functions. Actually
these are rational functions due to the presence of denominators related
to propagators. Second, there are also contributions quadratically dependent
on the coefficients of cubic in curvature terms in the potential.
Last there are contributions coming from operators quartic in curvature.
These terms contribute in a linearly dependent way in their coefficients. 
The full system of equations for the two conditions of vanishing of beta functions is rather too difficult
to solve for every unknown (there are rational, quadratic and linear type of equations).
However it is fairly easy to solve it for the coefficients of quartic
operators, even in the absence of cubic operators. And this is exactly,
what we did in this paper.

Let us summarize, what we need to make finite a super-renormalizable theory of gravity 
in $D=4$. First the coefficients in front of all four
terms of the type ${\cal R} \, \square^{\gamma-1}{\cal R}$ and ${\cal R} \, \square^{\gamma-2}{\cal R}$
must be set to zero to avoid running of the cosmological
constant. This corresponds to the minimal choice of the asymptotically polynomial form factor,
namely $p(z) = z^{\gamma+1}+O(z^{\gamma-2})$. 
Then it is optional or to put to zero all frontal coefficients
for terms cubic in curvature and with $2\gamma+2$ derivatives, either
to solve the linear equation for vanishing of the beta function for the 
Newton constant. This last option would express one coefficient
in terms of a linear dependence on all the others. 
By adjusting the parameters
of the theory in the potential to satisfy this choice, we get rid of perturbative
running of the Newton constant.
In order to kill the running of coupling constants in front of the operators quadratic
in curvature we also face with multiple choices. The minimal one is to
set to zero all cubic operators and to invoke only two terms, which
are quartic in curvature. The richer option is to take into account
all possible cubic and quartic operators at this order of $2\gamma+4$ 
derivatives of the metric. 
Later we may use two linear relations to make the two parameters
for the quartic operators dependent on all other parameters of terms quadratic,
cubic and quartic in curvature. (It is not known, if the same can be
achieved with only cubic operators.) By adjusting the two parameters
in the theory to satisfy this choice, we get rid of perturbative
running of the two quadratic coupling constants. Besides this there are
no other conditions to be imposed on those terms, if we demand perturbative
finiteness of the theory, provided that the conditions for one-loop
super-renormalizability are satisfied.

The minimal choice for a finite and unitary theory of quantum gravity in four
dimension may therefore consist of terms with $\gamma=3$ in the kinetic
part. 
For simplicity we may have only
two killers. The simplest Lagrangian 
may be the following,
\be
&& \hspace{-1cm}\mathcal{L}_{\rm fin} = - 2 \kappa_4^{-2}   \Big[ R  
+ R_{\mu \nu} \,  \frac{ e^{H(-\Box_{\Lambda})} -1}{\Box}   R^{\mu \nu} 
- \frac{1}{2} R \,  \frac{ e^{H(-\Box_{\Lambda})} -1}{\Box}   R
+s_{1}R^{2}\square R^{2}+s_{2}R_{\mu\nu}R^{\mu\nu}\square R_{\rho\sigma}R^{\rho\sigma} \Big] \,,
\nonumber 
\\
&& \hspace{-1cm} H(z) = \frac{1}{2} \left[ \Gamma \left(0,  p(z)^2 \right)+\gamma_E  + \log \left( p(z)^2 \right) \right]  ,
\ee
where $p(z) = z^{\gamma+1}=z^4$,
$s_{1}=-\frac{2\pi^2}{3}\omega_{2}(c_{1}+c_{2})$, $s_{2}=8\pi^2\omega_{2}c_{2}$ and 
$\omega_2  = e^{\gamma_E/2}/\Lambda^{2 \gamma+2}= e^{\gamma_E/2}/\Lambda^8$.
Here $c_{1}$ and $c_{2}$ are two constants independent on $\omega_{2}$,
that have to be determined from the calculation of the 
beta functions for the terms quadratic in the curvature.

More general Lagrangian can have a bunch of other terms (but still
finiteness of the theory can be obtained exactly in the same way):
\be
 && \hspace{-0.5cm}\mathcal{L}_{\rm fin} = - 2 \kappa_4^{-2}   \Big[ R  - \frac{\lambda}{2 \kappa_4^2} 
+ G_{\mu \nu} \,  \frac{ e^{H(-\Box_{\Lambda})} -1}{\Box}   R^{\mu \nu} 
+s_{1}R^{2}\square R^{2}
+s_{2}R_{\mu\nu}R^{\mu\nu}\square R_{\rho\sigma}R^{\rho\sigma} \Big] 
\nonumber \\
&& \hspace{1.57cm}
 - 2 \kappa_4^{-2}  \left[ {\sum}_i c_{i}^{(3)}\left({\cal R}^{3}\right)_{i}+ {\sum}_i c_{i}^{(4)}\left({\cal R}^{4}\right)_{i}+ 
{\sum}_i c_{i}^{(5)}\left({\cal R}^{5}\right)_{i} \right] \, . 
\ee
The last three terms have been written in a compact index-less notation. Note
that there are no covariant derivatives appearing there and that $c_{i}^{(3)}$, $c_{i}^{(4)}$ and $c_{i}^{(5)}$ are some constant coefficients.


\begin{thebibliography}{99}











\bibitem{modesto}
  L.~Modesto,
  Phys.\ Rev.\ D {\bf 86}, 044005 (2012)
  [arXiv:1107.2403 [hep-th]].

\bibitem{Briscese:2013lna}
  F.~Briscese, L.~Modesto and S.~Tsujikawa,
  Phys.\ Rev.\ D {\bf 89}, 024029 (2014)
  [arXiv:1308.1413 [hep-th]].





\bibitem{Tombo} 
E. T. Tomboulis
[hep-th/9702146v1].


\bibitem{Krasnikov}
  N.~V.~Krasnikov,
  Theor.\ Math.\ Phys.\  {\bf 73}, 1184 (1987)
  [Teor.\ Mat.\ Fiz.\  {\bf 73}, 235 (1987)].



\bibitem{BM} T. Biswas, E. Gerwick, T. Koivisto, A. Mazumdar,
Phys. Rev. Lett. 108, 031101 (2012).
[arXiv:1110.5249v2]; \\
 T. Biswas, T. Koivisto, A. Mazumdar [arXiv:1302.0532].

\bibitem{Chialva} 
  D.~Chialva and A.~Mazumdar,
  arXiv:1405.0513 [hep-th].


\bibitem{M2}	
  L.~Modesto,
  Astron. Rev. 8.2 (2013) 4-33
  [arXiv:1202.3151 [hep-th]];
  L.~Modesto,
  arXiv:1202.0008 [hep-th].




 \bibitem{M3}
  S.~Alexander, A.~Marciano and L.~Modesto,
  Phys.\ Rev.\ D {\bf 85}, 124030 (2012)
  [arXiv:1202.1824 [hep-th]].



\bibitem{M4}
  F.~Briscese, A.~Marciano, L.~Modesto and E.~N.~Saridakis,
  Phys.\ Rev.\ D {\bf 87}, 083507 (2013)
  [arXiv:1212.3611 [hep-th]].


\bibitem{Khoury:2006fg}
  J.~Khoury,
  Phys.\ Rev.\ D {\bf 76}, 123513 (2007)
  [hep-th/0612052].

\bibitem{modestoFinite}
  L.~Modesto,
  arXiv:1402.6795 [hep-th].

\bibitem{Mtheory}
  G.~Calcagni and L.~Modesto,
  arXiv:1404.2137 [hep-th].
  
  \bibitem{AnselmiW}
  D.~Anselmi,
  Phys.\ Rev.\ D {\bf 89}, 125024 (2014)
  [arXiv:1405.3110 [hep-th]].


\bibitem{NLsugra}
L. Modesto [arXiv:1206.2648 [hep-th]].




\bibitem{ModestoMoffatNico}
L. Modesto, J. W. Moffat, P. Nicolini,
Phys. Lett. B 695, 397-400 (2011)
[arXiv:1010.0680 [gr-qc]].



\bibitem{calcagniNL}
 G.~Calcagni, M.~Montobbio and G.~Nardelli,
 Phys.\ Lett.\ B {662}, 285 (2008)
 [arXiv:0712.2237 [hep-th]].


 \bibitem{E1} G. V. Efimov, ``Nonlocal Interactions" [in Russian], Nauka, Moscow (1977).

\bibitem{E2} G. V. Efimov, review paper [in Russian].

\bibitem{E3} V. A. Alebastrov, G. V. Efimov,
  V.~A.~Alebastrov and G.~V.~Efimov,
  Commun.\ Math.\ Phys.\  {\bf 31}, 1 (1973).


\bibitem{E4}
  V.~A.~Alebastrov and G.~V.~Efimov,
  Commun.\ Math.\ Phys.\  {\bf 38}, 11 (1974).




 \bibitem{E5}
  G.~V.~Efimov,
  Theor.\ Math.\ Phys.\  {\bf 128}, 1169 (2001)
  [Teor.\ Mat.\ Fiz.\  {\bf 128}, 395 (2001)].




\bibitem{Moffat1}
  J.~W.~Moffat,
  Phys.\ Rev.\ D {\bf 41}, 1177 (1990).





\bibitem{Moffat2}
  D.~Evens, J.~W.~Moffat, G.~Kleppe and R.~P.~Woodard,
  Phys.\ Rev.\ D {\bf 43}, 499 (1991).



\bibitem{Moffat3}
J.W. Moffat, Eur. Phys. J. Plus 126, 43 (2011)
[arXiv:1008.2482 [gr-qc]].





\bibitem{corni1} N.J. Cornish,
  N.~J.~Cornish,
  Mod.\ Phys.\ Lett.\ A {\bf 7}, 1895 (1992);
%
  N.~J.~Cornish,
  Mod.\ Phys.\ Lett.\ A {\bf 7}, 631 (1992).

\bibitem{MathStructure} L. Modesto and L. Rachwal, Mathematical Structure of Super-renormalizable 
Gravity, in preparation. 




\bibitem{odi}
  S.~'i.~Nojiri and S.~D.~Odintsov,
  Phys.\ Lett.\ B {\bf 659}, 821 (2008)
  [arXiv:0708.0924 [hep-th]];
  S.~Capozziello, E.~Elizalde, S.~'i.~Nojiri and S.~D.~Odintsov,
  Phys.\ Lett.\ B {\bf 671}, 193 (2009)
  [arXiv:0809.1535 [hep-th]];
  S.~'i.~Nojiri and S.~D.~Odintsov,
  Phys.\ Rept.\  {\bf 505}, 59 (2011)
  [arXiv:1011.0544 [gr-qc]].




\bibitem{Deser1}
  S.~Deser and R.~P.~Woodard,
  Phys.\ Rev.\ Lett.\  {\bf 99}, 111301 (2007)
  [arXiv:0706.2151 [astro-ph]].



\bibitem{Deser2}
  S.~Deser and R.~P.~Woodard,
  JCAP {\bf 1311}, 036 (2013)
  [arXiv:1307.6639 [astro-ph.CO]].



\bibitem{Modesto:2013jea}
  L.~Modesto and S.~Tsujikawa,
  Phys.\ Lett.\ B {\bf 727}, 48 (2013)
  [arXiv:1307.6968 [hep-th]];
  S.~Nesseris and S.~Tsujikawa,
  arXiv:1402.4613 [astro-ph.CO].


  \bibitem{Maggiore}
  M.~Jaccard, M.~Maggiore and E.~Mitsou,
  Phys.\ Rev.\ D {\bf 88}, 044033 (2013)
  [arXiv:1305.3034 [hep-th]];
  S.~Foffa, M.~Maggiore and E.~Mitsou,
  Phys.\ Lett.\ B {\bf 733}, 76 (2014)
  [arXiv:1311.3421 [hep-th]];
  Y.~Dirian, S.~Foffa, N.~Khosravi, M.~Kunz and M.~Maggiore,
  JCAP {\bf 1406}, 033 (2014)
  [arXiv:1403.6068 [astro-ph.CO]];
  G.~Cusin, J.~Fumagalli and M.~Maggiore,
  arXiv:1407.5580 [hep-th].



\bibitem{MazumdarIR} 
  A.~Conroy, T.~Koivisto, A.~Mazumdar and A.~Teimouri,
  arXiv:1406.4998 [hep-th].

\bibitem{BiswasSM} 
  T.~Biswas and N.~Okada,
  arXiv:1407.3331 [hep-ph].

\bibitem{Satz:2010uu} 
  A.~Satz, A.~Codello and F.~D.~Mazzitelli,
  Phys.\ Rev.\ D {\bf 82}, 084011 (2010)
  [arXiv:1006.3808 [hep-th]].

\bibitem{Reuter:2003yb} 
  M.~Reuter and F.~Saueressig,
  Fortsch.\ Phys.\  {\bf 52}, 650 (2004)
  [hep-th/0311056],
  Phys.\ Rev.\ D {\bf 66}, 125001 (2002)
  [hep-th/0206145].




\bibitem{shapiro}
M. Asorey, J.L. Lopez, I.L. Shapiro,
Intern. Journal of Mod. Phys. A12, 5711-5734 (1997)
[hep-th/9610006].



\bibitem{HigherDG0}
  F.~d.~O.~Salles and I.~L.~Shapiro,
  arXiv:1401.4583 [hep-th].






\bibitem{Anselmi:1991wb}
  D.~Anselmi,
  Phys.\ Rev.\ D {\bf 45}, 4473 (1992).

\bibitem{Anselmi:1992hv}
  D.~Anselmi,
  Phys.\ Rev.\ D {\bf 48}, 680 (1993).

\bibitem{Anselmi:1993cu}
  D.~Anselmi,
  Phys.\ Rev.\ D {\bf 48}, 5751 (1993)
  [hep-th/9307014].

\bibitem{Cnl1}
  G.~Calcagni, M.~Montobbio and G.~Nardelli,
  Phys.\ Lett.\ B {\bf 662}, 285 (2008)
  [arXiv:0712.2237 [hep-th]].



\bibitem{Cnl2}
  G.~Calcagni and G.~Nardelli,
  Phys.\ Rev.\ D {\bf 82}, 123518 (2010)
  [arXiv:1004.5144 [hep-th]].





 \bibitem{BambiMalaModesto2}
  C.~Bambi, D.~Malafarina and L.~Modesto,
  Phys.\ Rev.\ D {\bf 88}, 044009 (2013)
  [arXiv:1305.4790 [gr-qc]].
  
\bibitem{BambiMalaModesto} 
  C.~Bambi, D.~Malafarina and L.~Modesto,
  Eur.\ Phys.\ J.\ C {\bf 74}, 2767 (2014)
  [arXiv:1306.1668 [gr-qc]].


\bibitem{calcagnimodesto} 
  G.~Calcagni, L.~Modesto and P.~Nicolini,
  Eur.\ Phys.\ J.\ C in press
  [arXiv:1306.5332 [gr-qc]].


\bibitem{koshe1}
  A.~S.~Koshelev,
  Class.\ Quant.\ Grav.\  {\bf 30}, 155001 (2013)
  [arXiv:1302.2140 [astro-ph.CO]].


\bibitem{koshe2}
  T.~Biswas, A.~S.~Koshelev, A.~Mazumdar and S.~Y.~Vernov,
  JCAP {\bf 1208}, 024 (2012)
  [arXiv:1206.6374 [astro-ph.CO]].

\bibitem{koshe3}
  A.~S.~Koshelev and S.~Y.~Vernov,
  Phys.\ Part.\ Nucl.\  {\bf 43}, 666 (2012)
  [arXiv:1202.1289 [hep-th]].

\bibitem{koshe4}
  A.~S.~Koshelev,
  Rom.\ J.\ Phys.\  {\bf 57}, 894 (2012)
  [arXiv:1112.6410 [hep-th]].

\bibitem{V1}
  S.~Y.~Vernov,
  Phys.\ Part.\ Nucl.\  {\bf 43} (2012) 694
  [arXiv:1202.1172 [astro-ph.CO]].

\bibitem{kosheN}
  A.~S.~Koshelev and S.~Y.~Vernov,
  arXiv:1406.5887 [gr-qc].

  
  





\bibitem{HigherDG}
  A.~Accioly, A.~Azeredo and H.~Mukai,
  J.\ Math.\ Phys.\  {\bf 43}, 473 (2002);
  F.~d.~O.~Salles and I.~L.~Shapiro,
  arXiv:1401.4583 [hep-th].




 \bibitem{Stelle} 
  K.~S.~Stelle,
  Phys.\ Rev.\ D {\bf 16}, 953 (1977).




  \bibitem{Shapirobook} I. L. Buchbinder, Sergei D. Odintsov, I. L. Shapiro,
  ``Effective action in quantum gravity", IOP Publishing Ltd 1992.








\bibitem{GBV}
A. O. Barvinsky and Vilkovisky, Phys. Rep. 119, 1 (1985) 1-74.











\bibitem{VN} P. Van Nieuwenhuizen,
Nuclear Physics B 60 478-492 (1973).

\bibitem{AnselmiN} 
  D.~Anselmi,
  Phys.\ Rev.\ D {\bf 89}, 045004 (2014)
  [arXiv:1311.2704 [hep-th]].









\end{thebibliography}
\end{document}